\documentclass[USenglish,oneside,twocolumn]{article}

\usepackage{standalone}

\providecommand{\paperroot}{.}

\input{fig/common}
\providecommand{\netographDailyCapturesSum}{177868171}

\providecommand{\netographDailyCapturesTrancoHundredKSum}{107761991}

\providecommand{\satellitelogSizeOneTrust}{1492610}
\providecommand{\satellitelogSizeQuantCast}{5711528}

\providecommand{\shareInToplistDate}{Feb. 2021}

\providecommand{\OneTrustShareInTenK}{0.070068}
\providecommand{\QuantCastShareInTenK}{0.0277376}

\providecommand{\gvlDate}{Feb. 2021}
\providecommand{\gvlOneRevisions}{215}
\providecommand{\gvlTwoRevisions}{78}
\providecommand{\gvlAllRevisions}{293}
\providecommand{\gvlOneVendors}{602}
\providecommand{\gvlTwoVendors}{684}

\providecommand{\gvlAllPurposeChanges}{2103}

\providecommand{\googFalseTcfOld}{92}
\providecommand{\googFalseTcfNew}{91}
\providecommand{\googFalseTcfSwitch}{138}
\providecommand{\googTrueTcfOld}{378}
\providecommand{\googTrueTcfNew}{809}
\providecommand{\googTrueTcfSwitch}{664}

\providecommand{\qcTransitionOldCustomerLoss}{0.0656895}
\providecommand{\tcfOneCountInTrancoHundredK}{1539}
\providecommand{\tcfTwoCountInTrancoHundredK}{6726}
\providecommand{\tcfAllCountInTrancoHundredK}{7582}
\providecommand{\tcfAllCountInTrancoFiveK}{690}

\title{Privacy Preference Signals: Past, Present and Future}

\author[1]{Maximilian Hils}
\author[2]{Daniel W. Woods}
\author[3]{Rainer Böhme}

\affil[1]{University of Innsbruck, Austria, \newline E-mail: maximilian.hils@uibk.ac.at}
\affil[2]{University of Innsbruck, Austria, \newline E-mail: daniel.woods@uibk.ac.at}
\affil[3]{University of Innsbruck, Austria, \newline E-mail: rainer.boehme@uibk.ac.at}

\cclogo{\includegraphics{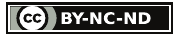}}

\journalname{Proceedings on Privacy Enhancing Technologies}
\DOI{10.2478/popets-2021-0069}
\startpage{249}
\received{2021-02-28}
\revised{2021-06-15}
\accepted{2021-06-16}

\journalyear{%
}
\journalvolume{2021}
\journalissue{4}

\begin{document}

\begin{abstract}{%
Privacy preference signals are digital representations of how users want their personal data to be processed. Such signals must be adopted by both the sender (users) and intended recipients (data processors).  Adoption represents a coordination problem that remains unsolved despite efforts dating back to the $1990$s.  Browsers implemented standards like the Platform for Privacy Preferences (P3P) and Do Not Track (DNT), but vendors profiting from personal data faced few incentives to receive and respect the expressed wishes of data subjects.  In the wake of recent privacy laws, a coalition of AdTech firms published the Transparency and Consent Framework (TCF), which defines an opt-in consent signal.  %
This paper integrates post-GDPR developments into the wider history of privacy preference signals.  
Our main contribution is a high-frequency longitudinal study describing how TCF signal gained dominance as of February $2021$.  We explore which factors correlate with adoption at the website level.  Both the number of third parties on a website and the presence of Google Ads are associated with higher adoption of TCF. Further, we show that vendors acted as early adopters of \tcftwo{} and provide two case-studies describing how Consent Management Providers shifted existing customers to \tcftwo{}.  %
We sketch ways forward for a pro-privacy signal.
}\end{abstract}
	
\maketitle

\vspace{-8ex}
\section{Introduction}
\vspace{-2ex}
Privacy preference signals are digital representations of how users want their personal data to be processed.  
These vary from a binary ``Do Not Track'' signal through to more complex expressions in cookie consent dialogues.  
Such signals are intended to influence how entities including websites and third parties process personal data. 
Web actors may collect privacy preferences in the hope of legitimizing data processing in the eyes of customers or to satisfy legal obligations.  

Efforts to standardize privacy preferences go back to at least P3P, which was presented as a prototype to US regulators in $1997$ and recommended as a standard by the World Wide Web Consortium (W3C) in 2002.  
It was adopted by around $20$k websites~\cite{leon2010token}, but was criticized by privacy advocates for not establishing consequences for false reporting of privacy practices~\cite{center2000pretty}.  
Another W3C working group was formed in 2011 to specify the Do Not Track HTTP extension but it was closed before completion, citing the lack of planned support among ``the ecosystem at large''~\cite{tpwg2019wg} as exemplified by the Interactive Advertising Bureau's withdrawal~\cite{iab2012dnt}.  
The first wave of privacy preference signals is completed by the opt-out cookies~\cite{dixon2007network} created by the Network Advertising Initiative (NAI) as part of a regulatory compromise with the Federal Trade Commission~\cite{united1998privacy}.  The NAI never published a specification, the opt-out only concerned a narrow definition of tracking, and very few vendors participated~\cite{dixon2007network}.

A second wave of privacy preference signals was prompted by the passage of privacy laws like the EU General Data Protection Regulation (GDPR) and the California Consumer Privacy Act (CCPA).  
For example, the GDPR establishes that an opt-in consent signal may constitute a legal basis for processing personal data providing the consent was ``freely given, specific, informed and unambiguous''.  %
These laws prompted research that has largely focused on the interfaces through which opt-in~\cite{utz2019uninformed, matte2020do, nouwens2020dark, machuletz2020} and opt-out~\cite{habib2020scavenger, oconnor2020unclear} signals are collected.
An ecosystem of actors has emerged to manage the collection of opt-in consent signals on behalf of websites~\cite{hils2020measuring}.  
Often these signals are collected and shared with a pay-for-membership ``Global Vendor List'', which has been termed the ``commodification of consent''~\cite{woods2020commodification}.

At this point, skeptics will rightly state that such signals exist in the world of \emph{soft privacy} with no technical guarantees about personal data flows and that we should instead focus on the technologies associated with \emph{hard privacy}. 
Such skepticism is compelling but should be qualified by the behavior of privacy advocates and AdTech firms.
Both sides invested resources in P3P and DNT working groups.  The latter posed a threat to AdTech business models as evidenced by the Interactive Advertising Bureau withdrawing from the working group after Microsoft announced it would be turned on by default~\cite{iab2012dnt}.  %
The power of these signals can also be seen in websites' dark patterns that nudge users towards expressing certain preferences~\cite{nouwens2020dark, habib2020scavenger, machuletz2020}.
Given the stakes have been further increased by sanctions associated with the GDPR and the CCPA, widespread adoption of a privacy preference signal would have privacy implications.

In terms of technical design, there is disagreement over who controls the interface by which users set privacy preferences.
In both P3P and DNT, the user expresses preferences to a user agent.  In contrast, user preferences are collected by embedding an interface in a web page in both of the approaches developed by AdTech industry bodies, namely the Interactive Advertising Bureau (IAB)~\cite{matte2020do} and the Network Advertising Initiative (NAI)~\cite{dixon2007network}.  
This bypasses browsers by making the signal backwards compatible with existing technology.  Turning to semantics, AdTech vendors proposed opt-in signals that could represent compliance, whereas privacy advocates proposed (global) opt-out signals that empower users.  In summary, these signals have a long history and also have privacy implications going forward.

\input{fig/privacy-standards-timeline}

This paper systematizes historical knowledge on privacy preference signals (the past), measures which signals have been adopted as of February 2021 (the present), and reflects on adoption strategies for a pro-privacy signal (the future).  We show a grim state of affairs for user control over privacy: P3P is obsolete, NAI's system still has only $\naiOptOutParticipants{}$ participating AdTech firms, and the reincarnation of Do Not Track---the Global Privacy Control---has been adopted by less than $10$ websites.  Meanwhile, the Interactive Advertising Bureau's \tcfone{} and \tcftwo{} have been adopted by thousands of websites. 
We then use high-frequency web measurements to build a longitudinal case-study of how adoption and \tcftwo{} migration varied over time, websites and AdTech vendors.  
Our contributions include:
\begin{itemize}
	\item \textbf{Systematize knowledge} about first wave (P3P, DNT, and NAI opt-out) and second  wave (TCF and GPC) privacy preference signals.
	\item \textbf{Measure present day adoption} and show that TCF adoption is roughly comparable to historical P3P adoption among websites, whereas an order of magnitude more AdTech vendors have adopted TCF than all other signals combined.
	\item \textbf{Test explanatory variables} for TCF adoption like website popularity, category, number of embedded third parties, and presence of Google Ads.  TCF adoption is higher among websites with closer ties to AdTech.
	\item \textbf{Longitudinal case-study} exploring \tcftwo{} migration strategies among the two most popular Consent Management Platforms, and how the new version changed the legal basis that individual AdTech vendors claim for tracking.  %
\end{itemize}

Section~\ref{section:background} describes the five privacy preference signals and Section~\ref{section:relatedwork} identifies related work measuring their adoption.  
This motivates our empirical measurements, which are described in Section~\ref{section:methods}.  Our results describing the present are contained in Section~\ref{section:results}.  
Section~\ref{section:discussion} discusses the past, present and future of privacy preferences. %
We conclude in Section~\ref{section:conclusion}.

\vspace{-2ex}
\section{Background} 
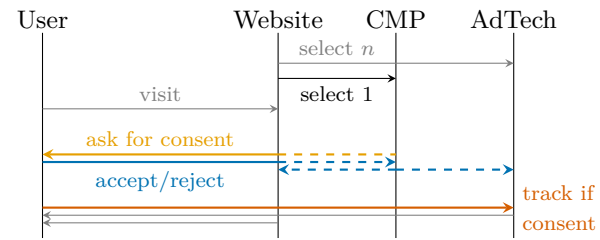
\begin{figure}[t!]

\tikzset{
	entity/.style = {
		above=-.2cm,
		execute at begin node={\strut}
	},
	signal/.style = {
		signal-color,
		thick
	},
	feedback/.style = {
		feedback-color, 
		thick
	},
	tracking/.style = {
		tracking-color,
		thick
	}
}

\textbf{Baseline: Personal data flow in web advertising}

\begin{tikzpicture}[x=15.5mm]
	\draw (-2,0) coordinate (U) node[entity] {User}       -- ++(0,-2);
	\draw ( 0,0) coordinate (W) node[entity] {Website}    -- ++(0,-2);
	\draw ( 2,0) coordinate (A) node[entity] {AdTech}     -- ++(0,-2);

	\small
	
	\draw[->,transform canvas={yshift=-0.4cm}] (W) -- node[above] {select $n$} (A);
	\draw[->,transform canvas={yshift=-0.8cm}] (U) -- node[above] {visit} (W);
	\draw[->,transform canvas={yshift=-1.2cm},tracking] (U) -- node[above=-.05cm,pos=.75] {personal data} (A) node[right] {track};

	\draw[<-,transform canvas={yshift=-1.6cm}] (U) -- node[below,pos=.75] {deliver ad} (A);
	\draw[<-,transform canvas={yshift=-1.8cm}] (U) -- node[below] {deliver} (W);

\end{tikzpicture}

~

\textbf{P3P privacy preference signal}

\begin{tikzpicture}[x=15.5mm]
	
	\draw (-2,0) coordinate (U) node[entity] {User}       -- ++(0,-1.8);
	\draw (-1,0) coordinate (B) node[entity] {Browser} -- ++(0,-1.8);
	\draw ( 0,0) coordinate (W) node[entity] {Website}    -- ++(0,-1.8);

	\small
	
	\draw[->,signal,transform canvas={yshift=-0.4cm}] (U) -- node[above] {conf. prefs.} (B);
	\draw[->,transform canvas={yshift=-0.6cm},gray] (U) -- node[above,pos=.75] {visit} (W);
	\draw[<-,transform canvas={yshift=-0.8cm}] (B) -- node[below] {priv. policy} (W);

	\draw[->,transform canvas={yshift=-1.2cm},feedback] (B) -- node[above] {notify} (U);
	
	\draw[->,transform canvas={yshift=-1.4cm},tracking] (U) -- node[below,pos=.25] {continue} (W) node[right] {track};

\end{tikzpicture}

~

\textbf{NAI opt-out privacy preference signal}

\begin{tikzpicture}[x=15.5mm]
	\draw (-2,0) coordinate (U) node[entity] {User}       -- ++(0,-2);
	\draw ( 0,0) coordinate (W) ++(0,-1) node[entity] {Website}    -- ++(0,-1);
	\draw ( 2,0) coordinate (A) node[entity] {AdTech}     -- ++(0,-2);

	\small
	\draw[->,transform canvas={yshift=-0.2cm},signal]   (U) -- node[above] {opt out} (A);
	\draw[<-,transform canvas={yshift=-0.3cm},feedback] (U) -- node[below] {confirm and set cookie} (A);
	\draw[->,transform canvas={yshift=-1.2cm},gray]     (W) -- node[above] {select $n$} (A);
	\draw[->,transform canvas={yshift=-1.4cm},gray]     (U) -- node[above] {visit} (W);
	\draw[->,transform canvas={yshift=-1.6cm},tracking] (U) -- (A) node[right] {\parbox{6em}{track if\\not opt-out}};
	\draw[<-,transform canvas={yshift=-1.7cm},gray]     (U) -- (A);
	\draw[<-,transform canvas={yshift=-1.8cm},gray]     (U) -- (W);

\end{tikzpicture}

~

\textbf{DNT/GPC privacy preference signal}

\begin{tikzpicture}[x=15.5mm]
	
	\draw (-2,0) coordinate (U) node[entity] {User}       -- ++(0,-1.7);
	\draw (-1,0) coordinate (B) node[entity] {Browser}    -- ++(0,-1.7);
	\draw ( 0,0) coordinate (W) node[entity] {Website}    -- ++(0,-1.7);
	\draw ( 2,0) coordinate (A) node[entity] {AdTech}     -- ++(0,-1.7);
	
	\small
	\draw[->,transform canvas={yshift=-0.4cm},signal]   (U) -- node[above] {opt out} (B);
	\draw[->,transform canvas={yshift=-0.975cm},signal, dashed] (B) -- (W);
	\draw[->,transform canvas={yshift=-1.275cm},signal, dashed] (B) -- (A);
	\draw[->,transform canvas={yshift=-0.8cm},gray]     (W) -- node[above] {select $n$} (A);
	\draw[->,transform canvas={yshift=-1.0cm},gray]     (U) -- node[above,pos=.75] {visit} (W);
	\draw[->,transform canvas={yshift=-1.3cm},tracking] (U) -- (A) node[right] {\parbox{6em}{track if\\not opt-out}};
	\draw[<-,transform canvas={yshift=-1.4cm},gray]     (U) -- (A);
	\draw[<-,transform canvas={yshift=-1.5cm},gray]     (U) -- (W);

\end{tikzpicture}

~

\textbf{TCF privacy preference signal}

\begin{tikzpicture}[x=15.5mm]
	
	\draw (-2,0) coordinate (U) node[entity] {User}       -- ++(0,-2.7);
	\draw ( 0,0) coordinate (W) node[entity] {Website}    -- ++(0,-2.7);
	\draw ( 1,0) coordinate (C) node[entity] {CMP}        -- ++(0,-2.7);
	\draw ( 2,0) coordinate (A) node[entity] {AdTech}     -- ++(0,-2.7);
	
	\small

	\draw[->,transform canvas={yshift=-0.4cm},gray]     (W) -- node[above,pos=.25] {select $n$} (A);
	\draw[->,transform canvas={yshift=-0.6cm}]          (W) -- node[below] {select 1} (C);
	\draw[->,transform canvas={yshift=-1.0cm},gray]     (U) -- node[above] {visit} (W);
	
	\draw[<-,transform canvas={yshift=-1.6cm},feedback] (U) -- node[above] {ask for consent} (W);
	\draw[- ,transform canvas={yshift=-1.6cm},feedback, dashed] (W) -- (C);
	\draw[- ,transform canvas={yshift=-1.7cm},signal]   (U) -- node[below] {accept/reject} (W);
	\draw[->,transform canvas={yshift=-1.7cm},signal, dashed] (W) -- (C);
	\draw[->,transform canvas={yshift=-1.8cm},signal, dashed] (C) -- (W);
	\draw[->,transform canvas={yshift=-1.8cm},signal, dashed] (C) -- (A);

	\draw[->,transform canvas={yshift=-2.3cm},tracking] (U) -- (A) node[right] {\parbox{2cm}{track if\\consent}};
	\draw[<-,transform canvas={yshift=-2.4cm},gray]      (U) -- (A);
	\draw[<-,transform canvas={yshift=-2.5cm},gray]      (U) -- (W);

\end{tikzpicture}

\caption{
	{\color{feedback-color} User prompt}, 
	{\color{signal-color}   privacy preference signals}, and
	{\color{tracking-color} personal data flows} when using each approach. %
}
\label{fig/flows}
\end{figure}

\vspace{-2ex}\label{section:background}
This section compares five privacy preference signals in terms of design properties and real-world adoption, which is summarized in Table~\ref{fig:standards-table}.  We selected these signals because they were the most widely adopted among the key stakeholders, namely browsers, AdTech vendors and websites.  We do not provide a background on the widespread online tracking that motivate privacy preference signals, such as cookies~\cite{acar2014theweb, englehardt2015cookies} and other tracking technologies~\cite{englehardt2016online, laperdrix2020browser, bujlow2017asurvey}. 
Similarly, we do not consider privacy preserving technologies unless they function to express privacy preferences, such as when browsers/add-ons collect user preferences and automate sending the signal.  We now turn to the five signals. Figure~\ref{fig:privacy-standards-timeline} provides and overview of the key events for each signal and Figure~\ref{fig/flows} provides a visual summary of the signal's flow.
 
\vspace{-1ex}
\subsection[Platform for Privacy Preferences (P3P)]{Platform\,for\,Privacy\,Preferences\,(P3P)}
\vspace{-1ex}
P3P is one of the earliest privacy preference signals proposed for the Web.
A demonstration of a P3P prototype was presented before the FTC in June 1997.  The W3C recommended the P3P 1.0 specification in $2002$, which describes an XML format to encode a human-readable privacy policy into a machine-readable specification stating the type, recipients and purposes of data collected.
Users can define individual privacy preferences, which browsers can cross-check against a website's self-reported P3P policy.
Each website's implementation could become arbitrarily complex with different policies for each web page and third-party cookie.%

P3P was adopted by, respectively, $588$ ($10\%$), $463$ ($8.34\%$), $2.3$k ($2.3\%$), and $33.1$k ($60\%$) of the sites in samples from 2003~\cite{byers2003automated}, $2007$~\cite{beatty2007p3p}, 2007~\cite{reay2013privacy}, and $2010$~\cite{leon2010token}.  The final sample~\cite{leon2010token} is not representative of the wider web because the majority of sites were discovered by the Privacy Finder search engine, which specifically aimed to identify web sites that respect a user's privacy.  However, the finding of $19\,820$ websites~\cite{leon2010token} implementing P3P in $2010$ serves as a reasonable lower bound in Table~\ref{fig:standards-table}.  The same study~\cite{leon2010token} found that $11$ ($15\%$) of a sample of AdTech vendors had a P3P privacy policy. 

Microsoft was the only browser developer to fully adopt P3P and stopped support in $2016$. Mozilla supported only some P3P features, but removed them by 2007. Other browsers shunned P3P and instead allowed users to set defaults like blocking all third party cookies~\cite{richmond2010loophole}.  P3P-specific browser extensions provide a more meaningful perspective on conscious user adoption than usage statistics for each browser.  For example, Privacy Bird, an add-on for Internet Explorer 5 and 6 that displays a website's P3P policy in an easy to understand language, was downloaded $20$k times in the first 6 months~\cite{cranor2002use}.

\vspace{-1ex}
\subsection{Network Advertising Initiative (NAI) Opt-Out} 
\vspace{-2ex}
AdTech vendors founded a self-regulatory body, the NAI, as a compromise following the Federal Trade Commission's (FTC) report on web privacy submitted to Congress in 1998~\cite{united1998privacy}.  The NAI established a system of opt-out cookies. Users can visit the NAI's website\footnote{\url{https://optout.networkadvertising.org/}} and set an opt-out cookie for each participating vendor to signal that the user does not want to be tracked by that firm. Critics~\cite{dixon2007network} note that the NAI's narrow definition of tracking would not cover many techniques observed in the wild~\cite{englehardt2016online, laperdrix2020browser, bujlow2017asurvey}.

The NAI provide a list of all participating vendors, which was just 4 in $2004$, %
$75$ in $2010$~\cite{leon2010token} and stands at $\naiOptOutParticipants{}$ participating vendors as of January $2021$. %
Websites and browsers do not need to adopt the NAI's system because it piggy-backs on existing browser cookie functionality. The NAI reported one million visits to the the opt-out page in $2006$~\cite{dixon2007network} but we cannot differentiate unique visitors.  Returning to browser extensions, there were at least $44.9$k users of the Targeted Advertising Cookie Opt-Out (TACO) add-on\footnote{\url{https://web.archive.org/web/20110920055245/https://addons.mozilla.org/en-us/firefox/addon/targeted-advertising-cookie-op/}}, which maintained an up to date list of opt-out cookies. %

\subsection{Do Not Track (DNT)}
Acknowledging the failure of P3P, the W3C created a working group in $2011$ to standardize the Do Not Track (DNT) mechanism~\cite{w3c2011dnt}.  DNT was less expressive than P3P.  Implementation involved browsers sending a \texttt{DNT:~1} header with each HTTP request to signal that their user did not wish to be tracked.  Stakeholders disagreed on whether DNT should default to on or off~\cite{iab2012dnt,wsj2012dnt}.  This opposition was part of the reason why the W3C working group was closed without success in $2019$~\cite{tpwg2019wg}.

DNT was implemented in browsers by Microsoft, Apple, Mozilla and eventually Google~\cite{chrome2012longer}.
Websites and third-party vendors could signal in an HTTP response header if they respected the user's DNT signal.  This signal was not exposed in any browser's user interface\footnote{\url{https://www.w3.org/TR/tracking-dnt/\#responding}} (outside of add-ons), which meant users were largely unaware of website adoption. Only 9 companies issued public statements regarding support of DNT~\cite{dntimplementors}. In $2011$, Mozilla reported DNT adoption by Firefox users to be at $17\%$ in the US and $11\%$ outside~\cite{fowler2013dnt}, although this oversamples privacy aware users. 

\subsection{Global Privacy Control (GPC)}  The unofficial GPC draft specification~\cite{gpc2020}, which was released in October $2020$, continues the work of DNT in extending HTTP requests with a single bit value.  %
Perhaps the most important change is re-framing \emph{Do Not Track} as a ``Do Not Sell'' and ``Object To Processing'' signal, which is closer to the language of the GDPR and the CCPA, which became effective in May $2018$ and January $2020$. This means GPC references (enforceable) laws, which DNT lacked.

As of February $2021$, Mozilla and the Brave browser are listed as publicly supporting GPC, but only Brave have implemented it. We do not provide any estimates for user size given it was released so recently.

\newcommand{\browserlogo}[1]{%
	\includegraphics[height=2ex]{\paperroot/fig/browsers/#1_256x256.png}%
}
\newcommand{\dead}{%
	\includegraphics[height=2ex]{\paperroot/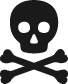}
}
\newcommand{\compat}{{\color{darkgray}--}}
\ifstandalone\begin{table}\else\begin{table*}\fi
	\centering
	\small
	\caption{Comparison of privacy preference signals} %
	\begin{tabular}{@{}lcccccc@{}} \toprule
		&P3P & NAI Opt-Out & DNT & GPC & \multicolumn{2}{c}{Transparency \& Consent Framework} \\ %
		\cmidrule(l){6-7}
		& & & & & TCF 1.x & TCF 2.x\\[-.5ex] \midrule
		\textbf{General} \\
		Convened by & W3C & AdTech \& FTC & \multicolumn{2}{c}{Privacy Advocates} & \multicolumn{2}{c}{AdTech} \\ 
		Legal Basis & none & self-regulat. & self-regulat. & CCPA & \multicolumn{2}{c}{GDPR} \\
		Standardized by & W3C & NAI & W3C & GPC & \multicolumn{2}{c}{IAB} \\ 

		\textbf{Design Properties}\\
		Implementation & Privacy Policy XML & Opt-Out Cookie & \multicolumn{2}{c}{HTTP Header} & \multicolumn{2}{c}{Consent Cookie from CMP} \\
		User Interface & UA indicator & central website & \multicolumn{2}{c}{UA setting or ext.} & \multicolumn{2}{c}{dialog on website} \\
		User Decision & configure prefs. & opt out & \multicolumn{2}{c}{turn on} &  \multicolumn{2}{c}{select allowed purposes} \\
		Decision Scope & all browsing & cookie lifetime & \multicolumn{2}{c}{all browsing} & \multicolumn{2}{c}{website until re-request} \\
		Vendor Decision & define policy & join NAI & \multicolumn{2}{c}{adopt standard} & \multicolumn{2}{c}{declare processing purposes} \\
		Website Decision & define policy & none & \multicolumn{2}{c}{adopt standard} & \hphantom{ii}pick vendors\hphantom{ii} & vendors\,+\,purposes  \\
		[1ex]
		\textbf{Adoption}\\
		Websites & $>20\text{k}$~\cite{byers2003automated, beatty2007p3p, leon2010token, reay2013privacy} & \compat & $>9$~\cite{dntimplementors} & ? &%
		$>\bignum{\tcfOneCountInTrancoHundredK}$*&%
		$>\bignum{\tcfTwoCountInTrancoHundredK}$*\\
		AdTech Vendors & $>11$~\cite{leon2010token} & $>75$~\cite{leon2010token} & $\approx 0$~\cite{dntimplementors} & ? 
		&$\gvlOneVendors$%
		&$\gvlTwoVendors$\\
		Browsers & \browserlogo{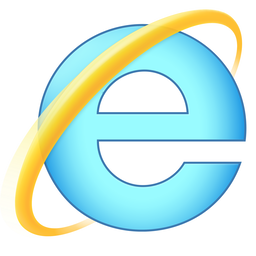} & \compat & \browserlogo{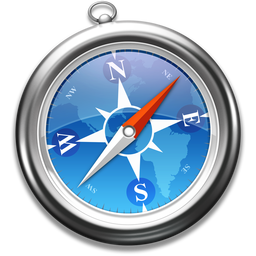}%
		\browserlogo{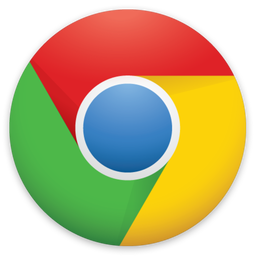}%
		\browserlogo{internet-explorer_9-11}%
		\browserlogo{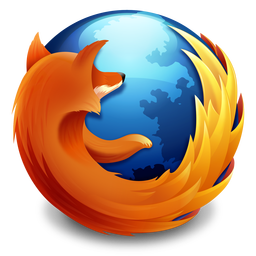} 
		& \browserlogo{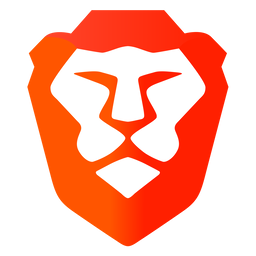}%
		& \compat & \compat  \\
		\bottomrule
		\multicolumn{7}{@{}l@{}}{ 
			?\,=\,unknown,\,%
			\compat{}\,=\,compat.\,with\,exist.\,tech.,\,%
			*\,=\,in\,Tranco\,100k,\,see\,Sec.\,\ref{section:results},\,%
			\browserlogo{safari_1-7}\,Safari
			\browserlogo{brave}\,Brave
			\browserlogo{chrome_12-48}\,Chrome
			\browserlogo{internet-explorer_9-11}\,Internet\,Explorer
			\browserlogo{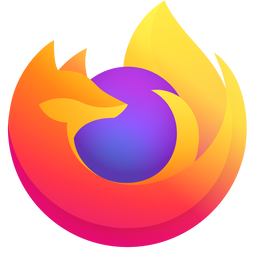}Firefox}%
	\end{tabular}
	\label{fig:standards-table}
\ifstandalone\end{table}\else\end{table*}\fi

\subsection{Transparency and Consent Framework (TCF)}

After the enactment of the GDPR, an advertising industry body (IAB Europe) formed a working group to develop the Transparency and Consent Framework (TCF), ``the only GDPR consent solution built by the industry for the industry''~\cite{iabeurope2020what}.
Participants predominantly representing private firms from the advertising and publishing industries co-developed the TCF, which defines the legal terms and data processing purposes that users consent to and the format by which consent signals are stored and exchanged between third parties.  A new version (\tcftwo{}) was introduced in $2020$. 

TCF is implemented by websites in the form of a consent dialog that does not require browser buy-in, much like NAI. It creates the role of \emph{Consent Management Providers (CMPs)}, who implement the framework on individual websites. CMPs are central to the TCF in providing an interface between website, user, and ad vendors. They provide websites with a (customizable) cookie prompt to embed, store users' choices as browser cookies, and provide an API for advertisers to access this information. We refer to Hils et al.~\cite[Fig. 2]{hils2020measuring} for a visual depiction of the ecosystem.

The IAB maintains a public list of CMPs, which lists 119 participating providers as of February 2021.\footnote{\url{https://iabeurope.eu/cmp-list/}}
A website wishing to implement the TCF independently must become a CMP, 
otherwise they can out-source this to an existing CMP. In reality, a handful of CMPs dominate the market~\cite{matte2020do}. The largest CMPs are OneTrust and Quantcast, which account for 37.4\% of all CMP implementations in the Tranco 100k (see Section \ref{section:results}).

To receive TCF consent signals from CMPs, AdTech vendors must register with the IAB and pay a yearly maintenance fee to join the \textit{Global Vendor List (GVL)}\footnote{\url{https://iabeurope.eu/vendor-list/}}. As of \gvlDate{}, \gvlTwoVendors{} companies are registered on this list. 
Most CMPs collect consent for the entire GVL by default, which means privacy preferences apply to the whole list~\cite{woods2020commodification}.

The specifications of \tcfone{} and \tcftwo{} both define a more complex signal than DNT/GPC. 
Under \tcfonex{}, users may affirmatively consent to any combination of five data processing purposes. They may also state individual preferences for each vendor on the GVL.
\tcftwo{} expands this model to ten purposes and two special features, increasing complexity even further.

In both TCF versions, users are prevented from expressing certain preferences. Vendors can claim that they have a legitimate interest in a specific purpose, which serves as their legal basis to process data even if the user clicks ``Reject all''. Starting with \tcftwo{}, some CMPs provide users with the additional option to object to this processing (GDPR asks for such functionality), but this needs to be done separately in a subdialog. As such, the ``Reject all'' button commonly does not actually express \emph{all} possible preferences. With \tcftwox{}, vendors can declare that their legal basis is flexible.  This means they would like to process data with the user's consent, but they can also perform (limited) processing based on a legitimate interest. 
As the only exception, \tcftwox{} removes the option for vendors to claim a legitimate interest in Purpose 1---``Store and/or access information on a device''---, possibly preempting an intervention by regulators. 
The policy changes between \tcfonex{} and \tcftwo{} motivate measuring the transition.

\section{Related Work} \label{section:relatedwork}
Section~\ref{subsection:practices} briefly describes the privacy practices employed by websites in order to motivate why privacy preferences matter.  Section~\ref{subsection:preferences} surveys research into  privacy preferences including the previous five signals.  Section~\ref{subsection:adoption} links the paper to the general question of \emph{why are technical standards adopted?}

\subsection{Privacy Practices} \label{subsection:practices}
Researchers consistently demonstrate privacy eroding techniques deployed in the wild~\cite{acar2014theweb, englehardt2015cookies, englehardt2016online, laperdrix2020browser, bujlow2017asurvey} motivated by online advertising business models~\cite{mayer2012third}.  Personal data is leaked via social networks~\cite{krishnamurthy2009leakage}, third-party web scripts~\cite{acar2020dataexfiltration}, apps~\cite{farooqi2020canarytrap}, software development kits~\cite{reyes2018think}, and organizational breaches~\cite{saleem2020sok}.  The scale of tracking motivate re-designing systems to provide privacy guarantees.  For example, multihoming can be used to defend against fingerprinting~\cite{henri2020multihoming} and trusted hardware can ensure compliance to stated privacy policies~\cite{mazmudar2020mitigator}.

Turning to so-called soft privacy, data processors are constrained by law and social norms.  These constraints are far from absolute.  For example, half of websites in a $2017$ sample violated laws implementing the EU Privacy Directive by installing cookies before collecting user consent~\cite{trevisan2019years}.  This is likely because organizations do not incur significant costs following data breaches and privacy violations in terms of either regulatory fines or lost shareholder value~\cite{WB2021-SoK}.  Nevertheless, firms' privacy practices are \emph{somewhat} impacted by data processors' self-declared privacy policies~\cite{shipp2020how, amos2020privacy, degeling2019,linden2020landscape} and even the privacy preferences expressed by users, to which we now turn.

\subsection{Privacy Preferences} \label{subsection:preferences}
Interviews~\cite{olson2005study} and surveys~\cite{ackerman1999privacy, weinshel2019oh} can use natural language to understand users' actual privacy preferences, which tend to contradict observed behavior~\cite{spiekerman2001eprivacy, barth2017privacy, gerber2018explaining}. 
Privacy languages aim to express preferences more precisely than natural language.  For example, APPEL encodes user preferences to be compared against P3P policies~\cite{cranor2003p3p}.  It could not express acceptable practices nor capture the realities of secondary sharing, which motivated XPref~\cite{agrawal2005xpref} and P2U~\cite{iyilade2014p2u}, respectively. Alternative languages focus on the usability for developers~\cite{yang2012language}, enabling audits~\cite{monir2015appl}, and providing explanations~\cite{kagal2008using}.  Privacy languages have been regularly surveyed by academics~\cite{kumaraguru2007survey, zhao2016privacylanguages, kasem2015security, morel2020} but unfortunately there has been little adoption in practice~\cite{zhao2016privacylanguages}.  This motivates our focus on signals deployed in the Web ecosystem.

In terms of the first wave of signals, measurements of DNT and NAI opt-out adoption relied on organizations disclosing private data sources like Firefox configurations~\cite{fowler2013dnt}, opt-out web page visits~\cite{dixon2007network}, or the NAI's membership~\cite{dixon2007network}.  P3P differed in that website adoption could be quantified via web scraping~\cite{byers2003automated, beatty2007p3p, cranor2008p3p, leon2010token, reay2013privacy, reay2009} often sampling via commercial website rankings.

Turning to the second wave, there are no GPC adoption studies because only a draft specification has been released so far.  The TCF ecosystem has been probed from a range of academic disciplines.   Legal methods are relevant to the semantic content of the signal.  For example, the purposes for collecting personal data standardized in the TCF may not be specific enough~\cite{matte2020purposes}.  

User interface research is important because the TCF does not standardize how the consent decision is presented to users, which is known to be influential~\cite{lai2006internet, boehme2010, adjerid2013sleights, machuletz2020}.  At least two studies have found that consent dialogues used to collect consent under the TCF contain design choices that nudge users towards providing consent~\cite{utz2019uninformed, nouwens2020dark}.

Web scraping studies have focused on implementation problems with TCF~\cite{matte2020purposes} or the ecosystem of consent management providers (CMP)~\cite{hils2020measuring}.  These studies provide measurements of TCF in passing.  For example, both studies measure TCF vendor registrations and their claimed purposes for processing data for \tcfonex{}~\cite[p.\,9]{hils2020measuring} and both \tcfonex{} and \tcftwo{}~\cite{matte2020purposes}.  The latter study measures aggregate  \tcftwo{} adoption, whereas we measure and visualise at the vendor level.  Matte et al.~\cite{matte2020do} show how \tcfonex{} adoption varies by top-level domain (TLD) and identify the most popular CMPs across the top 1k sites in five EU country code TLDs.   Hils et al.~\cite{hils2020measuring} use longitudinal measurements to show the market growth of six CMPs, highlighting how fast the ecosystem changes.

\subsection{Standards Adoption} \label{subsection:adoption}
We build on a body of work emphasizing the role of institutions in technical standards adoption.  For example, many vendors initially saw the TCP/IP protocols as a nuisance~\cite{leiner2009brief}. 
Leiner et al.~\cite{leiner2009brief} describe how a series of ``conferences, tutorials, design meetings and workshops'' were organized to educate a generation of vendors and engineers.  
The rest is history. 

The community was slow to turn to adoption questions like ``What Makes for a Successful Protocol?'', which was posed by RFC  5218 in $2008$.  Noting the qualitative nature of the resulting research, Nikkah et al.~\cite{nikkhah2017statistical} provide an illuminating statistical analysis of the association between technical features of $250$ RFCs and adoption success.  Analysing unchanging technical features cannot explain why it took two decades before IPv6 was widely adopted~\cite{czyz2014measuring, wang2018extending}.  Economic considerations like the scarcity of IPv4 addresses and the supply of compatible hardware can help explain \emph{when} standards are adopted~\cite{nikkah2016migrating}.

Thus, standards should be considered in the context of wider ecosystems governed by economic incentives.  For example, HTTPS adoption relies on X.509 certificate infrastructure that was ``in a sorry state'' in $2011$ with many websites relying on shared or invalid certificates~\cite{holz2011thessl}.  The situation was worse in the long tail likely because certificates are costly~\cite{ozment2006bootstrapping}.  Felt et al.~\cite{felt2017measuring} report on significant improvements in $2017$ and attribute improvements in the long tail to institutions like Let's Encrypt and publishing platforms---we show how similar economic considerations explain why TCF was adopted.

\subsection{Contribution}
Our main empirical contribution involves measuring the adoption of privacy preference signals among websites as of February $2021$.  Following the demise of P3P and DNT, the TCF has become dominant and the Global Privacy Control is still in its infancy.  We explore variables explaining which websites adopt TCF, and also longitudinally measure migration to a new version (\tcftwo{}).  %

This work differs from existing work by focusing exclusively on the adoption of privacy preference signals.  We largely ignore the actors~\cite{hils2020measuring, matte2020purposes} and interfaces~\cite{utz2019uninformed, nouwens2020dark, machuletz2020} harvesting such signals and instead focus on which factors (e.g.\,website type, popularity, and partners) are associated with TCF adoption.  Further, we are the first to systematize strands of research ranging from works in the late $1990$s to post-GDPR studies.  Finally, we provide the first results about migrating between versions of such signals using our the longitudinal methodology introduced in~\cite{hils2020measuring}.  Our previous work focuses on detecting specific CMPs, some of whom collect non-TCF signals exclusively or only collect TCF signals for a subset of customers.

\section{Methods} \label{section:methods}
We adopt a mixed approach\footnote{
	\textbf{Supplementary Material: \\\url{https://github.com/mhils/pets2021-privacy-preference-signals}}
	\label{fn:supplements}
} conducting both longitudinal high-frequency measurements to determine historic adoption of TCF and migration between versions, as well as a large-scale snapshot measurement to examine site-specific factors that may influence adoption. 
Section~\ref{methods:snapshot} describes our snapshot measurement of the Tranco 100k toplist. Section~\ref{methods:longitudinal} explains how we use the Netograph platform to conduct longitudinal high-frequency measurements.

\subsection{Snapshot Measurements} \label{methods:snapshot}
To measure the prevalence of TCF and its different versions on the web, we crawled the top 100k entries from the Tranco toplist, which aggregates the ranks from the lists provided by Alexa, Cisco Umbrella, Majestic, and Quantcast~\cite{pochat2019}. Our automated browser crawls were performed in February 2021 using a Tranco toplist from January 2020\footnote{Available at \url{https://tranco-list.eu/list/K8JW}}. We used this older toplist dated shortly before publishers transitioned to \tcftwox{} in order to avoid survivorship bias in our observations.  Picking a later toplist would over-sample websites created post-2020 who are certain to adopt \tcftwo{} and de facto avoid a migration decision. Our toplist and a current Tranco toplist (Tranco id KGNW from Feb. 19th 2021) overlap by 76.5\%.

We first converted the Tranco list of domains to a list of URLs that can be crawled. For each \textit{domain}, we attempted to establish a TLS and a TCP connection with www.\textit{domain} and \textit{domain} on port 443 and 80, respectively. 
This was repeated three times over a week to catch temporary service disruptions. We then picked a configuration that was reachable at least once, preferring TLS over TCP and secondly www.\textit{domain} over \textit{domain} to construct our crawl URL. An error in the TLS certificate verification was treated as unreachable. We used \textit{http://domain} as a fallback if no connections were successful.

Our crawling infrastructure was set up in a European university network. Websites were opened using Google Chrome on Linux with its current default user agent,\footref{fn:supplements} a desktop resolution of 1024$\times$800, and \texttt{en-US} as the preferred browser language. All other settings were set to their defaults: third party cookies are allowed, the DNT and GPC HTTP headers are not set. The low desktop resolution and all other settings were chosen to match that of our longitudinal measurements described below. Crawls are automated using custom browser instrumentation based on the Chrome DevTools Protocol. Unsuccessful crawls were retried twice within a week.

For every capture, we collected the following data points using custom browser instrumentation. First, HTTP headers are stored for all requests and responses. Second, connection-related metadata such as IP addresses and TLS certificate chains are logged. Third, for every domain in a capture, its relation to the main page, all cookies, IndexedDB, LocalStorage, SessionStorage and WebSQL records are saved. Fourth, we store the browser's DOM tree and record a full-page screenshot (including scrolling).

\begin{table}
	\centering
	\small
	\pgfkeys{/pgf/fpu=true}
	\providecommand{\none}{{\color{gray}--}}
	\providecommand{\ngAll}{%
		\bignum{int(\netographDailyCapturesSum / 100000)/10}M%
	}
	\providecommand{\ngTop}{%
		\bignum{int(\netographDailyCapturesTrancoHundredKSum / 1000000)/10}M%
	}
	\providecommand{\ngQC}{\bignum{int(\satellitelogSizeQuantCast / 100000)/10}M}
	\providecommand{\ngOT}{\bignum{int(\satellitelogSizeOneTrust  / 100000)/10}M}
	\providecommand{\ngQCOT}{\bignum{int((\satellitelogSizeQuantCast+\satellitelogSizeOneTrust)  / 100000)/10}M}
	\providecommand{\tranco}{$100$k}
	\providecommand{\gvl}{$\gvlAllRevisions$}
	
	\providecommand{\row}[7]{%
		#1 & #2 & #3 & #4 & #7 \\
	}
	\caption{Data sources for figures.}
	\label{fig:data-sources}
	\setlength{\tabcolsep}{5pt}
	\begin{tabular}{@{}rlrrr@{}} \toprule
		Figure & Approach & Data Source & $N$ & CMP\\
		\midrule
\row{\ref{fig:tcf-bytoplist}, \ref{fig:tcf-by-tp}}      {Snapshot\,(Feb.\,'21)}     {Tranco\,Toplist} {\tranco} {Snapshot Crawls}              {100k}  {all}   
\row{\ref{fig:tcf-by-category}}      {Snapshot\,(Feb.\,'21)}     {Tranco\,Toplist} {$10$k} {Snapshot Crawls}              {10k}  {all}   
\row{\ref{fig:google-impact}, \ref{fig:tcf-share-timeline}}      {Longitudinal} {Netograph}      {\ngQCOT} {Longitudinal Crawls}          {100k}  {QC/OT} 
\row{\ref{fig:qc-transition}}      {Longitudinal} {Netograph}      {\ngQC}   {Longitudinal Crawls}          {\none} {QC}    
\row{\ref{fig:ot-transition}}      {Longitudinal} {Netograph}      {\ngOT}   {Longitudinal Crawls}          {\none} {OT}    
\row{\ref{fig:gvl-transition}--\ref{fig/gvl-transition-storage}}     {Diff.\,of\,vendor\,list}  {IAB}            {\gvl}    {Global Vendor List Revisions} {\none} {\none} 
		\bottomrule
	\end{tabular}
\end{table}

\subsubsection{TCF Adoption}
We automatically detect whether crawled websites implement the TCF. To do this, we wait for the website's DOMContentLoaded event to fire, then wait another ten seconds, and then inject JavaScript code into the execution context of the root document. This approach for CMP detection was already validated by Matte et al.~\cite{matte2020do} with more aggressive timeouts. As each CMP must implement a \texttt{\_\_cmp()} function for \tcfonex{} and \texttt{\_\_tcfapi()} function for \tcftwox{}, we check for the presence of these functions to determine if TCF is being used. We additionally checked for other signs of TCF (such as the presence of \texttt{\_\_tcfapiLocator} or \texttt{\_\_cmpLocator}), but this search did not turn up any new results.
For every TCF API we find, we issue a \texttt{ping} command to learn more about the implementation. In the case of \tcftwox{}, the \texttt{PingReturn} object (as specified by the TCF) is expected to contain the CMP's identifier (as assigned by the IAB) as well as the CMP/GVL/TCF versions in use. 
We also considered that a CMP may masquerade as a different CMP here. We correlated the reported CMP ids with contacted domains and did not find any evidence of misrepresentation.

The adoption of TCF is naturally higher on some types of websites, such as those who typically display paid advertisements. %
To quantify this, we divided the Tranco 10k toplist into categories with the help of Symantec Rulespace~\cite{rulespace}, a categorization database already used in related work by Sanchez-Rola et al.~\cite{sanchez2019}.
We limit our analysis to the Tranco 10k as a non-negligible share of websites ($11.7$\%) in the top 100k is not categorized, compared to only $2.4\%$ for the top 10k websites.
We note that recent work has shown that most categorization services are not fit for detecting specialized content or content-blocking~\cite{vallina2020}, but this does not significantly affect our coarse classification of popular domains.

To determine the number of third parties present on each website, 
we normalized all requested URLs to their effective second-level domain using Mozilla's Public Suffix List~\cite{publicsuffix}. This list contains all suffixes under which internet users can directly register names, including non-standard ``TLDs'' such as \texttt{blogspot.com}. We note that this approach does not account for recent obfuscation techniques such as CNAME cloaking \cite{dimova2021cname}.

We also examined the fraction of websites that appear to be collecting data versus those showing a cookie prompt. To determine a lower bound, we took all third-party domains that were included on at least 1.000 websites in the Tranco 100k (158 domains) and manually removed shared resources such as content delivery networks which may not constitute tracking (12 domains). We then determined for each website if any of the remaining 146 third parties were embedded. For example, we exclude \texttt{s3.amazonaws.com} as this domain is commonly used to serve static assets and not for tracking. In contrast, almost all remaining domains clearly belong to ad companies. We include both lists in the supplementary material.\footref{fn:supplements}

Finally, we estimated the prevalence of non-TCF cookie notices or consent prompts in our snapshot measurements using a simple back-of-the-envelope heuristic. For every capture, we scan the stored copy of the browser's final DOM tree for the occurrence of the phrase ``cookie''. The resulting estimates only indicate orders of magnitude, which is acceptable given they are not core to any of our results.  Rather they are intended to provide context, such as showing government websites are significantly less likely to present a cookie notice than our other categorizations (see Figure~\ref{fig:tcf-by-category}). In a manual inspection of 50 randomly picked domains with and 50 domains without ``cookie'' in their DOM tree, we found five domains that had a ``Cookie Notice'' link in their footer (but no dialog) and no false negatives (which yields a 5\% error rate overall). Again, this part of our analysis is not as rigorous as our other measurements and is only intended to provide context in Figure~\ref{fig:tcf-by-category}.

\subsection{Longitudinal Measurements} \label{methods:longitudinal}
To measure the adoption and transition between TCF versions longitudinally, we analyze automated browser crawls recorded by the Netograph web measurement platform.\footnote{\url{https://netograph.io/}} Netograph continuously ingests a live feed of social media posts, extracts all URLs, and visits them from crawlers located in EU and US data centers. For brevity, we refer to \cite{hils2020measuring} for a discussion of the validity and reliability of this measurement method. Most importantly, HTTP message contents are not retained due to storage constraints, but a large amount of metadata is stored, such as the HTTP headers of every request.

Relying on metadata in our longitudinal data means we have to measure TCF adoption using CMP-specific indicators.  Instead of building quick and dirty heuristics for over $90$ CMPs, we focus our efforts on creating a set of reliable indicators for two of the leading providers in the consent management market, Quantcast and OneTrust, which are embedded on \bignum{int((\QuantCastShareInTenK + \OneTrustShareInTenK)*1000)/10}\% of websites in the Tranco 10k (\shareInToplistDate{}). We manually analyzed their respective dialog implementations and identified distinct HTTP requests that indicate the use of specific TCF versions\footref{fn:supplements}. For Quantcast, we detected the use of TCF for all implementations dating back to May 2018. For OneTrust, we identified the use of \tcfone{} or \tcftwo{} in their Cookie Consent SDK launched at the end of 2019 (\texttt{otSDKStub.js}).

From Netograph's \bignum{int(\netographDailyCapturesSum / 1000000)} million captures in the social media dataset, we obtained all \bignum{int(\satellitelogSizeQuantCast / 100000)/10} million captures that include a Quantcast consent dialog and all \bignum{int(\satellitelogSizeOneTrust / 100000)/10} million captures that include a OneTrust consent dialog. We grouped captures by their effective second-level domain to not overcount repeated measurements with varying subdomains. Due to Netograph's sampling strategy, less popular domains may not be observed for a several days. We account for this by explicitly marking the period between the last \tcfone{} and the first \tcftwo{} measurement as an (unobserved) transition phase.

\subsubsection{Measuring Vendor Adoption}
To track the adoption of \tcftwo{} by AdTech vendors, we downloaded all previously published lists of vendors registered as participating in the TCF from the IAB and verified their accuracy using the Internet Wayback Machine. These lists include each vendor's declared purposes for processing personal data. As of \gvlDate{}, there are \gvlOneRevisions{} revisions of this list for \tcfone{} and \gvlTwoRevisions{} revisions for \tcftwo{}. We then inspected these previous versions for longitudinal changes and measured every instance when an AdTech vendor joins, leaves, or switches to \tcftwo{}. 
While \tcftwo{} is not backwards compatible from a publisher's point of view, a vendor that has declared support for \tcftwo{} may still accept \tcfone{} consent strings from publishers.

\begin{figure}
	\begin{tikzpicture}[x=.31cm,y=.0025cm]
		
		\newcommand{\tcfShare}[9]{
			\begin{scope}[draw=darkgray]
				\path (#1,0)++(-0.5,0) coordinate (x);
				\draw[fill=other-color]     (x) rectangle ++(1,#7+#8) ++(-1,0) coordinate (x);
				\draw[fill=liveramp-color]     (x) rectangle ++(1,#4) ++(-1,0) coordinate (x);
				\draw[fill=google-color]     (x) rectangle ++(1,#6) ++(-1,0) coordinate (x);
				\draw[fill=sourcepoint-color]     (x) rectangle ++(1,#5) ++(-1,0) coordinate (x);
				\draw[fill=quantcast-color] (x) rectangle ++(1,#2   ) ++(-1,0) coordinate (x);
				\draw[fill=onetrust-color]  (x) rectangle ++(1,#3   ) ++(-1,0) coordinate (x);
			\end{scope}
		}
	
		\input{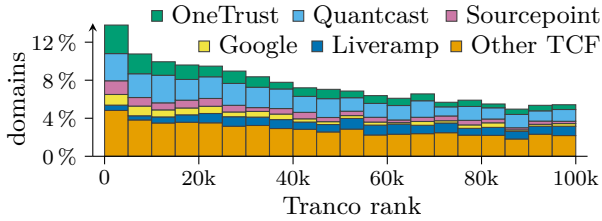}

		\draw[->] (-1,0) -- (-1,700);
		\draw (-1,0) -- (20,0);
		
		\node at (10,-4.5ex) {Tranco rank};
		\node[rotate=90] at (-4.25,300) {domains};
		
		\newcommand{\legend}[3]{
			\draw[draw=darkgray,fill=#2] (#1) rectangle ++(4pt,4pt) ++(0,-2pt) node[right] {\strut#3};
		}
		\legend{1.5,710}{onetrust-color}{OneTrust}
		\legend{7.5,710}{quantcast-color}{Quantcast}
		\legend{14,710}{sourcepoint-color}{Sourcepoint}
		\legend{3.5,550}{google-color}{Google}
		\legend{8.3,550}{liveramp-color}{Liveramp}
		\legend{14.2,550}{other-color}{Other TCF}
		
		\small
		\foreach \y in {0,200,400,600} {
			\draw (-1,\y) -- ++(-3pt,0) node[left] {$\bignum{\y/5000*100}\,\%$};
		}
	
		\foreach \x/\lbl in {0/0,4/$20$k,8/$40$k,12/$60$k,16/$80$k,20/$100$k} {
			\draw (\x-.5,0) -- ++(0,-3pt) node[below, inner sep=2pt] {\lbl};
		}
	\end{tikzpicture}
	\caption{Share of websites in the Tranco 100k that use a CMP. OneTrust and Quantcast are the most popular providers, followed by Sourcepoint, Google, and Liveramp.}
	\label{fig:tcf-bytoplist}
\end{figure}

\begin{figure}
	\centering
	\begin{tikzpicture}[x=8cm,y=.42cm]
		
		\definecolor{other-prompt-color}{HTML}{BBBBBB}		
		
		\newcommand{\tcfDist}[5]{
			\pgfmathsetmacro\total{#2 + #3 + #4 + #5}
			\pgfmathsetmacro{\nothirdparties}{1-#1/\total}
			\pgfmathsetmacro{\none}{#2/\total}
			\pgfmathsetmacro{\misc}{\none + (#3/\total)}
			\pgfmathsetmacro{\tcfA}{\misc + (#4/\total)}

			\draw[draw=lightgray, fill=none] (0,-.5) rectangle (\none,.5);
			\draw[draw=gray,fill=other-prompt-color] (\none,-.5) rectangle (\misc,.5);
			\draw[draw=darkgray,fill=v1-color] (\misc,-.5) rectangle (\tcfA,.5);
			\draw[draw=darkgray,fill=v2-new-color] (\tcfA,-.5) rectangle (1,.5);
			
			\draw[fill=black,draw=black] (\nothirdparties,-.5)++(0,2pt*0.86) -- ++(1pt,-2pt*0.8) -- ++(-2pt,0) -- cycle;		

		}
		\foreach \i/\x/\lbl in {
			5/news/News \& Entertainment,
			4/shopping/Shopping,
			3/technology/Technology,
			2/business/Business,
			1/education/Education,
			0/government/Government%
		} {
			\begin{scope}[yshift=\i*.5cm]
				\input{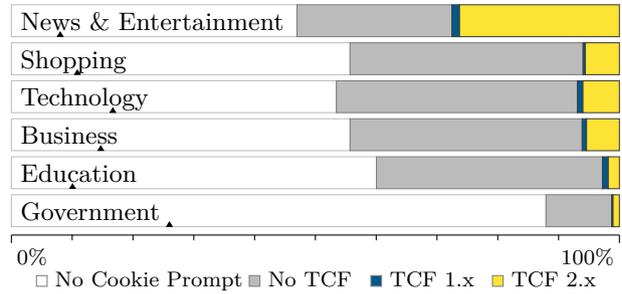}
				\node[right] at (0,-.2ex) {\strut\lbl};
			\end{scope}
		}
	
		\footnotesize
		\begin{scope}[yshift=-.32cm]
			\draw (0,0) -- ++(1,0);
			\draw (0,0) -- ++(0,-3pt) node[below right=-1pt] {0\%};
			\draw (1,0) -- ++(0,-3pt) node[below left=-1pt] {100\%};
			\foreach \i in {.1,.2,.3,.4,.5,.6,.7,.8,.9} {
				\draw (\i,0) -- ++(0,-2pt);	
			}
		\end{scope}

		\small
		\newcommand{\legend}[3]{
			\draw[draw=gray,fill=#2] (#1,-2.2)++(-2pt,-2pt) rectangle ++(4pt,4pt) ++(0,-2pt) node[right] {\strut#3};
		}
		\legend{.05}{white}{No Cookie Prompt}
		\legend{.4}{other-prompt-color}{No TCF}
		\legend{.6}{v1-color}{TCF 1.x}
		\legend{.8}{v2-new-color}{TCF 2.x}
	\end{tikzpicture}
	\caption[]{
		Share of websites in the Tranco 10k with a (TCF)~cookie prompt. For reference, \tikz[baseline=-1mm]{
			\draw[fill=black,draw=black] (0,0) -- ++(1pt,-2pt*0.8) -- ++(-2pt,0) -- cycle;
		} marks the share of websites which do not embed popular third parties.
	}
	\label{fig:tcf-by-category}
\end{figure}

\section{Results} \label{section:results}
Section~\ref{results:adoption} focuses on the relationship between website characteristics and TCF adoption mainly using snapshot measurements.  Section~\ref{results:tcf2} explores how vendors and websites migrated to \tcftwo{} using our longitudinal approach. Table~\ref{fig:data-sources} maps each figure to the approach, data source, and covered CMPs. We provide the underlying data in the supplementary material.\footref{fn:supplements}

\subsection{TCF Adoption}  \label{results:adoption}
We first explore how TCF adoption varies by the popularity and category of website.  Figure~\ref{fig:tcf-bytoplist} shows that TCF is more prevalent among popular websites (e.g\,the Tranco 5k) and that adoption is relatively consistent through the Tranco 100k.  Websites embedding OneTrust comprise a greater fraction of TCF implementations for more popular sites (Tranco 20k), whereas Quantcast embeds are more evenly distributed.  Quantcast's free self-service solution may be better suited to less popular sites than OneTrust's, which requires an interaction with a sales associate.  By offering a free and usable solution, Quantcast is playing a similar role to Let's Encrypt with HTTPS adoption~\cite{felt2017measuring}.

Figure~\ref{fig:tcf-by-category} shows that TCF adoption in the Tranco top 10k is highest among websites classified as News $\&$ Entertainment and is lowest among Government websites.  The grey bars provide a relatively coarse indication (see the previous section) of what percentage of each category displays a cookie prompts.  Few Government websites display prompts, which helps to explain the low TCF adoption.  Almost half of all cookie prompts on News $\&$ Entertainment sites implement TCF, whereas this fraction is less than $15\%$ for each of the other five even though the first five categories have a similar fraction of websites showing cookie prompts.  This motivates exploring alternative explanations.

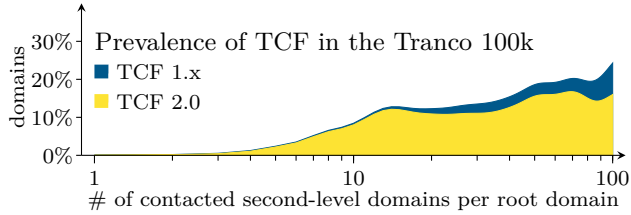
\begin{figure}
	\centering
	\begin{tikzpicture}[x=97pt,y=5cm]

	\begin{scope}[every path/.style={v1-color, fill}]
		\input \paperroot/data/tcf2-transition/tcf_adoption_by_third_parties/tcf1_stacked.tex -- (2,0) -- cycle;
	\end{scope}
	\begin{scope}[every path/.style={v2-new-color, fill}]
		\input \paperroot/data/tcf2-transition/tcf_adoption_by_third_parties/tcf2.tex -- (2,0) -- cycle;
	\end{scope}

	\draw[->] (-.05,0) -- (-.05,.4);
	\draw[->] (-.05,0) -- (2.05,0);
	
	\node[right, inner xsep=0] at (0,1.5cm) {Prevalence of TCF in the Tranco 100k};	
	
	\small
	\draw[fill=v1-color, draw=none] (0, 1.4cm-1*\baselineskip) rectangle ++(1.5ex,1.5ex) -- ++(0,-.95ex) node[right] {\strut\tcfonex{}};
	\draw[fill=v2-new-color, draw=none] (0, 1.4cm-2*\baselineskip) rectangle ++(1.5ex,1.5ex) -- ++(0,-.95ex) node[right] {\strut\tcftwo{}};
	
	\node at (1, -.12) {\# of contacted second-level domains per root domain};
	
	\node[rotate=90] at (-.3,.18) {domains};
	
	\small
	
	\foreach \y in {0, 10, 20, 30}{
		\draw (-.05,\y/100) -- ++(-2pt,0) node[left, inner sep=1pt] {$\y\%$};
	};

	\foreach \e in {0,1} {
		\foreach \i in { 2, ..., 10 } {
			\pgfmathparse{log10((10^\e) * \i)}

			\draw (\pgfmathresult,0) -- ++(0,-2pt);
		};
		\pgfmathsetmacro{\hosts}{int(10^(\e+1)}
		\draw (\e+1,0) -- ++(0,-3pt) node[below] {$\hosts$};
	}
	\draw (0,0) -- ++(0,-3pt) node[below] {$1$};
		
	\end{tikzpicture}
	\caption{Adoption of the TCF increases significantly for websites that embed a large number of third parties.}
	\label{fig:tcf-by-tp}
\end{figure}
\begin{figure}
	\centering
	\begin{tikzpicture}[y=.0025cm, x=.7cm]
		
		\input{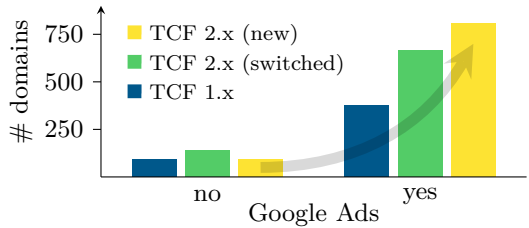}
		
		\newcommand{\drawbar}[3]{
			\path[fill=#3] (#1,0)++(-0.42,0)  rectangle ++(.84,#2);
		}
		
		\drawbar{1}{\googFalseTcfOld}{v1-color}
		\drawbar{2}{\googFalseTcfSwitch}{v2-switch-color}
		\drawbar{3}{\googFalseTcfNew}{v2-new-color}

		\drawbar{5}{\googTrueTcfOld}{v1-color}
		\drawbar{6}{\googTrueTcfSwitch}{v2-switch-color}	
		\drawbar{7}{\googTrueTcfNew}{v2-new-color}

		\begin{scope}[transparency group, opacity=0.15]
			\draw (3, 50) edge[bend right=30, ->, black, line width=1ex] (7, 700);
		\end{scope}
		
		\draw[->] (0,0) -- ++(0,900);
		\draw (0,0) -- (8,0);
		
		\foreach \y in {250,500,750} {
			\draw (0,\y) -- ++ (-3pt,0) node[left] {\y};
		}

		\node at (2,-.25cm) {no};
		\node at (6,-.25cm) {yes};
		\node at (4,-.5cm) {Google Ads};
		
		\node[rotate=90] at (-1.5, 500) {\# domains};
		
		\small
		\foreach \i/\x/\lbl in {
			1/v2-new/TCF 2.x (new),
			2/v2-switch/TCF 2.x (switched),
			3/v1/TCF 1.x
		} {
			\draw[fill=\x-color, draw=none] (0.5, 2.2cm-\i*\baselineskip) rectangle ++(1.5ex,1.5ex) -- ++(0,-.95ex) node[right] {\strut \lbl};
		}
		
	\end{tikzpicture}
	\caption{Google did not participate in \tcfonex{} and only joined \tcftwo{}. Their partners' websites were far more likely to adopt \tcftwox{} but not \tcfonex{}. 
	}
	\label{fig:google-impact}
\end{figure}

We explored whether web relationships can help explain varying adoption rates.    Figure~\ref{fig:tcf-by-tp} shows that TCF adoption increases with the number of embedded third parties.  
This result could be caused by third parties influencing partner websites to adopt TCF, but it could also be mere correlation.   Websites with business models based on personal data may be \emph{both} more likely to embed many third parties and also more likely to adopt the TCF.  

Causality could be probed via a natural experiment in which websites were randomly assigned a partner that exerts influence.  It can be argued the decision of Google to join \tcftwo{} but not \tcfonex{} provides such an opportunity.  By comparing the relative adoption of \tcfonex{} and \tcftwo{} among websites which embed Google with those who do not, we can isolate the effect on TCF adoption of partnering with Google.  If partnering with Google influences websites' decisions, we would expect a higher fraction of such websites to adopt \tcftwo{} but not \tcfonex{} as compared to the same fraction among non-partners.  
Indeed, Figure~\ref{fig:google-impact} shows that for websites supporting \tcftwox{} and not using Google Ads, \pgfmathparse{\googFalseTcfSwitch/(\googFalseTcfSwitch + \googFalseTcfNew)*100}\pgfmathprintnumber[precision=0]{\pgfmathresult}\%
had already joined \tcfonex{}, whereas this applies to only 
\pgfmathparse{\googTrueTcfSwitch/(\googTrueTcfSwitch + \googTrueTcfNew)*100}\pgfmathprintnumber[precision=0]{\pgfmathresult}\%
of the websites using Google Ads.
We cannot tell whether the influence is active (e.g.\,vendor X only contracts with TCF websites) or passive (e.g.\,website Y finds it easier to adopt the same standard as their partners).

To shed more light on these relationships, we run logistic regressions with \tcftwo{} adoption as the dependent variable.  For each website, we have the following explanatory variables: a binary dummy for the presence of Google ads $\beta_1$ (from Figure~\ref{fig:google-impact}), log of the number of embedded third parties $\beta_2$\footnote{We count the first party domain so that $\beta_2 \geq 0$.} (from Figure~\ref{fig:tcf-by-tp}), and the website category (from Figure~\ref{fig:tcf-by-category}).  We include a full regression table in the Appendix (Table~\ref{fig/regression}).  

As we would expect from the figures, the first regression shows $\beta_1$ and $\beta_2$ have a positive relationship with adoption:
\begin{equation} \label{model:security}
y \approx -4.6^{\ast\ast\ast} +  0.15^{\ast\ast\ast} \beta_1  + 0.77^{\ast\ast\ast} \beta_2 
\end{equation}
and both effects are statistically significant at the $p=0.01$ level.  This means each variable adds additional explanatory power.  

Model~2 adds a fixed effect for each website category and this boosts the Pseduo-$R^2$ from 0.08 to 0.13 relative to Model~1.  The coefficient for News \& Entertainment is positive and highly significant.  The high adoption rate among such websites exceeds what could be explained by  $\beta_1$ and $\beta_2$ alone. 

Finally, Model~3 explores the interaction effect between $\beta_1$ and $\beta_2$.  The sign of  $\beta_1 * \beta_2$ means that the relationships are sub-additive---the increased likelihood of adoption from increasing both variables is less than the sum of increasing each variable independently.  Although these regressions have shown that website category and web relationships help explain \tcftwo{} adoption rates, the Pseduo-$R^2$ shows a lot of the variance remains unexplained.  This could be down to our relatively crude statistical design aiming to directly link variables to organisation-level outcomes.  A recent systematization of knowledge~\cite{WB2021-SoK} highlights similar difficulties explaining cybersecurity outcomes via manifest variables and suggests  latent variables inferred via reflexive indicators represent a better way forward.

\input{\paperroot/fig/tcf-transition-common.tex}
\begin{figure}
	\hfill %
	\begin{tikzpicture}[y=2.1cm/9000,x=2.35cm]
		
	\node[right] at (18.05,2.3cm) {Quantcast Configurations};	
	
	\foreach \i/\x/\lbl in {
		1/v2-new/TCF 2.x (new),
		2/v2-switch/TCF 2.x (switched from 1.x),
		3/transition/Transition, 
		4/v1/TCF 1.x
	} {
		\begin{scope}[every path/.style={\x-color, fill}]
			\input{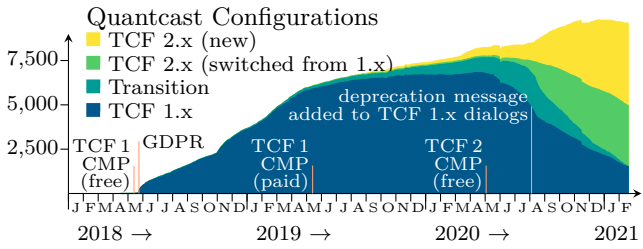}
		\end{scope}
		\small
		\draw[fill=\x-color, draw=none] (18.1, 2.25cm-\i*0.8*\baselineskip) rectangle ++(1.5ex,1.5ex) -- ++(0,-.95ex) node[right] {\strut \lbl};
	}
	
	\timeaxis

		\draw[->] (18,0) -- (18,9000);
		\foreach \y in {2500,5000,7500}{
			\draw (18,\y) -- node [left=2pt, inner sep=0] {\small\pgfmathprintnumber[fixed]{\y}} ++(-3pt, 0);
		};
		
		\draw[eventline] (18.3671,0) -- ++(0,0.8\baselineskip) node[event,left,align=right] {TCF\,1\\[-.5ex]CMP\\[-.5ex](free)};  %
		\draw[eventline] (18.3945,0) -- ++(0,1.5\baselineskip) node[event,right] {GDPR};  %
		
		\draw[eventline] (19.3671,0) -- ++(0,0.8\baselineskip) node[event,white,left,align=right] {TCF\,1\\[-.5ex]CMP\\[-.5ex](paid)}; %
		 
		\draw[eventline] (20.3388,0) -- ++(0,0.8\baselineskip) node[event,white,left,align=right] {TCF\,2\\[-.5ex]CMP\\[-.5ex](free)};
		
		\draw[eventline] (20.5929,0) -- ++(0,2.5\baselineskip) node[event,white,left,align=right] {deprecation message\\[-.5ex]added to \tcfonex{} dialogs}; %

	\end{tikzpicture}
	\caption{TCF Adoption by Quantcast customers. Note that the y-axis differs from OneTrust; Quantcast started with a significantly larger number of \tcfonex{} customers.}
	\label{fig:qc-transition}
\end{figure}

\subsection{\tcftwo{} Migration} \label{results:tcf2}
The release of \tcftwo{} provides an opportunity to observe how actively both vendors and websites adopt these signals.

\providecommand{\transitionPlot}[2]{
	\node[right] at (18.05,2.3cm) {#1};	
	
	\foreach \i/\x/\lbl in {
		1/v2-new/TCF 2.x (new),
		2/v2-switch/TCF 2.x (switched from 1.x),
		3/transition/Transition, 
		4/v1/TCF 1.x
	} {
		\begin{scope}[every path/.style={\x-color, fill}]
			\input{\paperroot/data/tcf2-transition/#2/stacked/\x.tex}
		\end{scope}
		\small
		\draw[fill=\x-color, draw=none] (18.1, 2.25cm-\i*0.8*\baselineskip) rectangle ++(1.5ex,1.5ex) -- ++(0,-.95ex) node[right] {\strut \lbl};
	}
	
	\timeaxis
}

\begin{figure}
	\hfill %
	\begin{tikzpicture}[y=2.1cm/1500,x=\timeaxisx]
		
	\node[right] at (18.05,2.3cm) {OneTrust Configurations};	
	
	\foreach \i/\x/\lbl in {
		1/v2-new/TCF 2.x (new),
		2/v2-switch/TCF 2.x (switched from 1.x),
		3/transition/Transition, 
		4/v1/TCF 1.x
	} {
		\begin{scope}[every path/.style={\x-color, fill}]
			\input{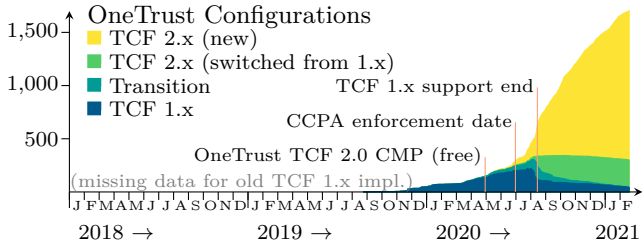}
		\end{scope}
		\small
		\draw[fill=\x-color, draw=none] (18.1, 2.25cm-\i*0.8*\baselineskip) rectangle ++(1.5ex,1.5ex) -- ++(0,-.95ex) node[right] {\strut \lbl};
	}
	
	\timeaxis

		\draw[->] (18,0) -- (18,1700);
		\foreach \y in {500,1000,1500}{
			\draw (18,\y) -- node [left, inner sep=0] {\small\pgfmathprintnumber[fixed]{\y}\;} ++(-3pt, 0);
		};

		\node[above right, inner sep=0] at (18,0) {\scriptsize\color{gray}(missing data for old \tcfonex{} impl.)};
		\draw[eventline] (20.3279,0) -- ++(0,1\baselineskip) node[event,left] {OneTrust \tcftwo{} CMP (free)\strut};
		\draw[eventline] (20.4973,0) -- ++(0,2\baselineskip) node[event,left] {CCPA enforcement date\strut};
		\draw[eventline] (20.6202,0) -- ++(0,3\baselineskip) node[event,left] {\tcfonex{} support end\strut};

	\end{tikzpicture}
	\caption{TCF Adoption by OneTrust customers. Most \tcfonex{} customers switched to \tcftwox{} around August 2020. Since July 2020, OneTrust gained a large number of new customers which directly started using \tcftwox{}. \textit{Transition} marks the unobserved interval during which a switch from \tcfonex{} to 2.x occurred. }
	\label{fig:ot-transition}
\end{figure}

\subsubsection{Websites} Quantcast have the most customers embedding TCF, claim to be a driving force behind its development, and launched a new free \tcftwo{} product in May $2020$.  Yet Figure~\ref{fig:qc-transition} shows how a large share of their customers had not adopted the new version when \tcfonex{} support by the IAB ended on August 15th. Approaching the IAB's deadline, Quantcast went as far as embedding a prominent deprecation notice visible to all website visitors into its \tcfone{} consent dialogs (see Figure \ref{fig:qc-attention-site-owner}).
Quantcast lost customers while enforcing the switch over, which can be seen in the fall ($\bignum{int(100*\qcTransitionOldCustomerLoss)}\%$) in old customers who had implemented \tcfone{} from the start of August to end of September.  Quantcast's total customers continue to grow due to new customers who directly adopt \tcftwo{} (the yellow fraction), but the fall in old customers can be seen in the decreasing total of the green and blue lines in Figure \ref{fig:qc-transition}.

\begin{figure}
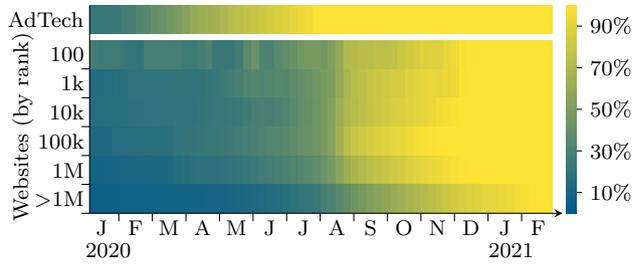

	\begin{tikzpicture}[y=.38cm,x=5.3cm]
		
		\def\xmin{20.0109}
		\def\xmax{21.22}
	
		\newcommand{\heatmap}[3]{
			\pgfmathparse{#1 >= \xmin}
			\ifnum \pgfmathresult > 0
				\pgfmathparse{min(100,100*#3*(1/0.8))}
				\fill
					[v2-new-color!\pgfmathresult!v1-color]
					(#1,7-#2) rectangle ++(7.5/365,1); %
			\fi
		}
		\input{\paperroot/data/tcf2-transition/heatmap.tex}

		\newcommand{\heatmapGvl}[2]{
			\pgfmathparse{#1 >= \xmin}
			\ifnum \pgfmathresult > 0
			\pgfmathparse{min(100,100*#2*(1/0.8))}
			\fill
				[v2-new-color!\pgfmathresult!v1-color]
				(#1,6.2) rectangle ++(7.5/365,1); %
				\fi
		}
		\input{\paperroot/data/gvl/heatmap.tex}
		\path (\xmin,6.2)++(-4pt, .5) node[inner sep=0, left=-2pt] {\small AdTech};

		\draw (\xmin,0) -- (\xmin,7.2);
		\draw[->] (\xmin,0) -- (21.185,0);

		\small
		
		\foreach \y/\lbl in {
			5/100,
			4/1k,
			3/10k,
			2/100k,
			1/1M,
			0/>1M
		}{
			\draw (\xmin,\y) -- ++(-3pt, 0);
			\path (\xmin,\y)++(-4pt, .5) node[left=-5pt] {\lbl};
		};
	
		\foreach \x/\lbl in {
			1/J
			,2/F
			,3/M
			,4/A
			,5/M
			,6/J
			,7/J
			,8/A
			,9/S
			,10/O
			,11/N
			,12/D
			,13/J
		} {
			\draw (20+\x/12,0) -- ++(0,-3pt);
			\node at (20-1/24+\x/12,-6pt) {\strut\lbl};
		}
	
		\node at (20-1/24+14/12,-6pt) {\strut F};
	
		\node[right, inner sep=0] at (20, -14pt) {2020};
		\node[right, inner sep=0] at (21, -14pt) {2021};

		\path (\xmin,3)++(-25pt,0) node[rotate=90, inner sep=0] {Websites (by rank)};
	
		\fill [top color=v2-new-color, bottom color=v1-color] (\xmax,0) rectangle ++(-4pt,7.2);
		\foreach \p in {10,30,50,70,90}{
			\draw (\xmax,\p/100*7.2) -- ++(2pt,0) node[right] {\p\%};
		}

	\end{tikzpicture}
	\caption{Share of websites in each segment of the Tranco toplist that use the TCF and have upgraded to version 2.x.}
	\label{fig:tcf-share-timeline}
\end{figure}

In contrast, OneTrust lost very few customers in transition, which can be seen in the bright green area in Figure~\ref{fig:ot-transition}.  OneTrust acquired many new customers from June $2020$ and the majority of these immediately adopted \tcftwo{}.  As a result, OneTrust had a higher fraction of customer implementing \tcftwo{} than Quantcast by the end of September $2020$ even though Quantcast pursued a more assertive transition strategy. However, Quantcast remain comfortably ahead of OneTrust in terms of number of websites embedding TCF (although OneTrust also implements a significant number of non-TCF dialogs~\cite{hils2020measuring}).

Returning to the role of top list position, Figure~\ref{fig:tcf-share-timeline} shows that websites in the Tranco top $100$ began experimenting with \tcftwo{} migration in the first half of $2020$.  The experimentation can be seen in how migration went down at various points.  The majority had permanently transitioned by July $2020$.  This suggests the CMP's announcement about ending support for \tcfone{} were sufficient to lead to migration for popular websites.  However, the less popular websites were far less responsive.

\subsubsection{Vendors}

\begin{figure}
	\hfill %
	\begin{tikzpicture}[y=2.1cm/700,x=\timeaxisx]
		\node[right] at (18.05,2.3cm) {Vendors on the Global Vendor List};	
		
		\foreach \i/\x/\lbl in {
			1/v2-new/TCF 2.x (new),
			2/v2-switch/TCF 2.x (switched),
			3/v1/TCF 1.x
		} {
			\begin{scope}[every path/.style={\x-color, fill}]
				\input{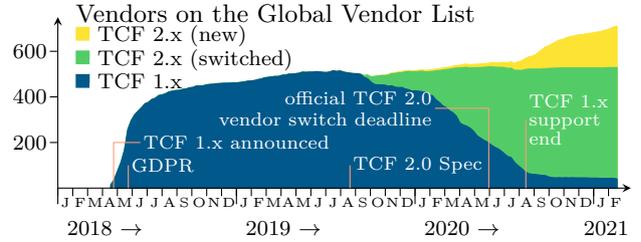};
			\end{scope}
			\small
			\draw[fill=\x-color, draw=none] (18.1, 2.25cm-\i*0.8*\baselineskip) rectangle ++(1.5ex,1.5ex) -- ++(0,-.95ex) node[right] {\strut \lbl};
		}
		
		\timeaxis
		
		\draw[->] (18,0) -- (18,750);
		\foreach \y in {200,400,600}{
			\draw (18,\y) -- node [left, inner sep=0] {\small\pgfmathprintnumber[fixed]{\y}\;} ++(-3pt, 0);
		};

		\draw[eventline] (18.3123,0) |- ++(0.15,200) node[event,right,white] {\tcfonex{} announced};
		\draw[eventline] (18.3945,0) -- ++(0,100) node[event,right,white] {GDPR};  %
		\draw[eventline] (19.6356,0) -- ++(0,100) node[event,right,white] {\tcftwo{} Spec};
		\draw[eventline] (20.4126,0) |- ++(-.3,350) node[event,left,white,align=right] {official \tcftwo{} \\vendor switch deadline};
		\draw[eventline] (20.6202,0) -- ++(0,300) node[event,right,align=left,white] {\tcfonex{}\\[-.5ex]support\\[-.25ex]end};
		
	\end{tikzpicture}
	\caption{TCF Adoption by ad-tech vendors.}
	\label{fig:gvl-transition}
\end{figure}

The majority of early adopters were vendors rather than websites.  By the start of $2020$, more vendors ($84$) had switched to \tcftwo{} than there were websites ($48$) embedding either version of TCF using OneTrust's Consent SDK.  Figure~\ref{fig:gvl-transition} shows vendors appear to follow an $S$-growth pattern with slow uptake, a relatively small window in which the majority adopt, and a stubborn tail.  The number of vendors implementing each version of TCF was relatively consistent through to September 2020, which suggests the upgraded TCF was not a major draw for vendors unlike for websites embedding Google Ads (see Figure~\ref{fig:google-impact}).  The growth rate increased from September 2020 for reasons we do not know, but this is much smaller than the post-GDPR growth.

Comparing time to adoption and migration between vendors and websites speaks to the question of which constituency is driving TCF adoption. Figure~\ref{fig:gvl-transition} shows most vendors had already adopted \tcfone{} by the time GDPR came into effect, whereas OneTrust had no TCF product and only a fraction of Quantcast's $2020$ customers were implementing TCF.  The same pattern holds for \tcftwo{} migration.  This is consistent with vendors providing an incentive for partner websites towards adoption.  While we cannot claim causality, this evidence at least makes it unlikely that websites pushed vendors towards adoption.

\begin{figure}
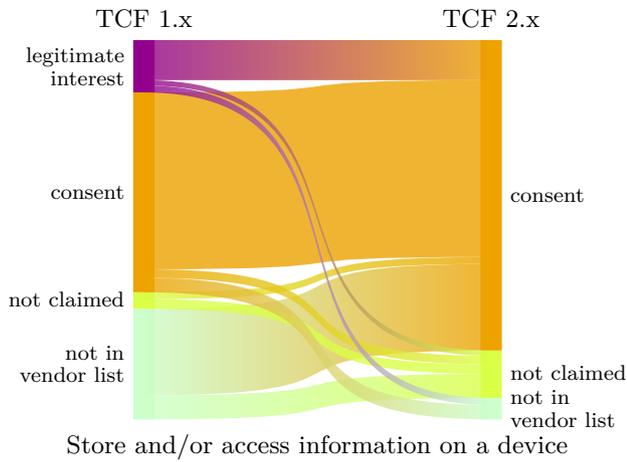

	\centering
	\begin{tikzpicture}[y=.2pt,x=130pt]
		
		\input{\paperroot/fig/gvl-transition-common.tex}
		\input{\paperroot/data/gvl/purpose_sankey/p1_p1.tex}
		
		\path (0,\sankeyCount)++(0,2ex) node[align=center] {\tcfonex{}};
		\path (1,\sankeyCount)++(0,2ex) node[align=center] {\tcftwox{}};
		
		\path (.5,-2.5ex) node[align=center] {Store and/or access information on a device};
	\end{tikzpicture}
	\caption{Removing the option to claim legitimate interest for purpose 1 of the TCF (see Section~\ref{section:background}) led more vendors to collect consent for accessing information such as advertising identifiers under \tcftwox{}. New vendors that did not adopt \tcfonex{} (\textit{not in vendor list}) mostly seek consent as well.
	}
 	\label{fig/gvl-transition-storage}
\end{figure}

\subsubsection{Implications}  Thus far we have focused on adoption and migration without considering the details or privacy implications of the switch.  We illustrate the need for future work by measuring the effect of migrating to \tcftwo{} on the legal basis by which vendors claimed the right to process personal data.  We recount some of the background from Section~\ref{section:background}.  Both versions of TCF define purposes for processing personal data. \pagebreak For each purpose, vendors implementing \tcfonex{} can declare either; they do not use personal data for that purpose, need to first obtain consent before doing so, or claim they have a legitimate interest in doing so (which users cannot dispute).

The IAB removed the option to claim a legitimate interest in storing and/or accessing information on a device under \tcftwox.  Figure~\ref{fig/gvl-transition-storage} shows how this shifted the majority of vendors who were previously claiming legitimate interest towards asking for consent.  This highlights how standards setters can influence how privacy preferences are communicated at scale by removing the legally questionable options.

\begin{figure}
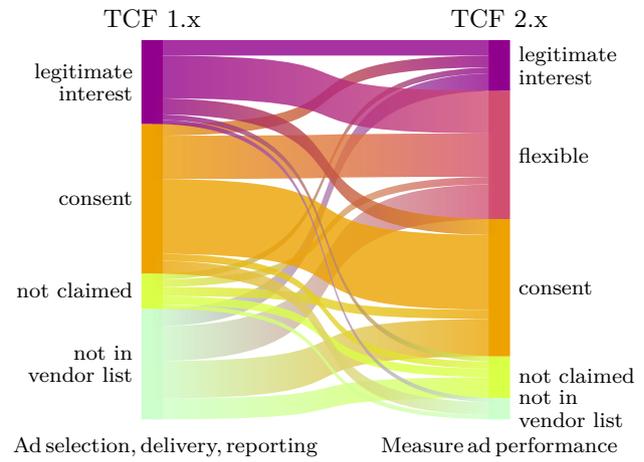

	\centering
	\begin{tikzpicture}[y=.2pt,x=130pt]
		
		\input{\paperroot/fig/gvl-transition-common.tex}
		\input{\paperroot/data/gvl/purpose_sankey/p3_p7.tex}
		
		\path (0,\sankeyCount)++(0,2ex) node[align=center] {\tcfonex{}};
		\path (1,\sankeyCount)++(0,2ex) node[align=center] {\tcftwox{}};
		
		\path (0,-2.5ex)++(5pt,0) node[align=center, inner xsep=0] {\small Ad\,selection,\,delivery,\,reporting};
		\path (1,-2.5ex) node[align=center, inner xsep=0] {\small Measure\,ad\,performance};
	\end{tikzpicture}
	\caption{In migrating from \tcfonex{} to \tcftwox{}, a large portion of vendors now can claim to be flexible regarding the legal basis; i.e. they will perform the processing based on consent or a legitimate interest.}
 	\label{fig:adperf}
\end{figure}

Updated standards can also add complexity that makes analyzing impacts difficult to evaluate.  For example, the purpose ``ad selection, delivery and reporting'' was renamed and split into multiple purposes in \tcftwox{}.  Additionally, vendors had the additional option to declare that they are flexible regarding the legal basis; they can perform the processing based on consent or a legitimate interest.  Figure \ref{fig:adperf} shows how this led to a decrease in both the number claiming legitimate interest and also the number collecting consent, which means its unclear whether users lost or gained control under the new standard.  These results show just one way in which the design of standards impacts user privacy.

\newpage
\section{Discussion} \label{section:discussion}
This section discusses the past, present (as established in the previous section) and future of privacy preference signals.

\subsection{Past} Mark Twain's quip that ``history doesn't repeat itself, but it often rhymes'' is also true of privacy preference signals, and identifying these rhymes helps to reason about the present and future.  For example, Table~\ref{fig:standards-table} shows that signals proposed by AdTech (NAI and TCF) collect user preferences via a web page, whereas the signals proposed by privacy advocates are collected by a browser.   As a result, browsers immediately support AdTech signals and could only stop them by actively preventing web content rendering, meanwhile AdTech vendors must actively make the decision to support P3P, DNT and GPC.  Consequently,  standards developed by AdTech industry bodies have been adopted by browsers by default, whereas AdTech vendors can delay adoption and thus undermine the standard.

Privacy preference signals also vary in terms of the signal's scope, permanence, and how decision volume scales with web usage.  Table~\ref{fig:standards-table} highlights how privacy preferences are collected in a single interaction under P3P and DNT/GPC and the browser assumes that this decision applies to the entire Web.  Consequently, the user makes a single decision that has long-term signaling implications.  In contrast, the NAI's opt-out cookies only apply to specific forms of tracking~\cite{dixon2007network} and only last until the user loses the cookie or the vendor sets a new one.

Scope and permanence are even narrower under the TCF, which contains asymmetries based on the preferences expressed.  The decision not to provide consent\footnote{Notably, the TCF framework does not even mention the possibility a user can ``revoke'' a decision~\cite{iabeurope2020what}.} only applies to a specific website and only last until the website re-requests consent, whereas positive consent signals may apply to multiple websites~\cite{woods2020commodification, matte2020do} and re-requests are less frequent. Table~\ref{fig:standards-table} shows history repeating itself in that privacy advocates support a signal that imposes a low decision load on users (P3P, DNT and GPC), whereas AdTech vendors support impermanent signals with a narrow application that force a decision burden on users (NAI and TCF).

Turning to the forum in which signals were designed, we have seen a movement away from development via consensus-based working groups committed to open standards.  Initially all parties met in working groups coordinated by the W3C but the clashing political objectives led to splintering.  For example, the Digital Advertising Alliance withdrew from the DNT working group in $2012$ citing the lack of progress~\cite{inside2013digital}. 

The second wave of privacy preference signals were developed outside of open, consensus-based groups.  TCF was developed via a working group listing $139$ participating organizations~\cite{iabtechlab2020global} for which the Interactive Advertising Bureau controlled membership. The resulting TCF signal is closed in that both websites and vendors need the IAB's permission to implement it, although this authority is delegated to consent management providers. 
GPC is developed more openly, but lists only $17$ supporting organizations with no formal forum to coordinate development.  For comparison, the P3P 1.0 specification lists participants from $56$ organizations, the DNT working group contained $110$ members~\cite{inside2013digital}, and the NAI for a long time only included ``a fraction of the industry''~\cite{dixon2007network} and now counts $\naiMembers{}$ members.  

In retreating to less consensus-based processes, the Global Privacy Control and the Interactive Advertising Bureau follow (in more than just initials) the governance model of the Internet Advisory Board, which was created in $1984$ to incorporate stakeholders beyond Vint Cerf's ``kitchen cabinet''~\cite[p.~51]{russell2006rough}:
\begin{quote}
	``The IAB cannot be characterized as a democracy, since nobody voted and the Board only let in the people they wanted \dots{} Democracy, with its competing factions and its political compromises, was not an appropriate political model for the IAB or the Internet.''
\end{quote}
The same could be true of privacy standards given over $10$ years was spent drafting P3P and DNT at the W3C.  It should be noted that the Internet's IAB later moved towards more open governance by creating and transferring power to the IETF~\cite{russell2006rough}.  It seems unlikely AdTech's IAB will voluntarily follow suit, which raises the question of regulatory involvement.

The history of privacy preference signals is intertwined with regulation.  Do Not Track began as a letter to congress and was re-invigorated by the FTC chairman going off script to mention it years later~\cite{soghoian2011thehistory}. The NAI's opt-out cookies resulted from an agreement with the FTC to self-regulate~\cite{dixon2007network}.  The IAB created the TCF in response to the GDPR, and GPC quotes ``Do Not Sell'' directly from the CCPA. However, none of these signals are mandated by law, which means they could become de-facto standards by achieving widespread adoption.

A final lesson from history is that for all the willingness of browser developers to attend working groups, they are reluctant to support privacy preference signals if doing so risks impacting user experience.  For example, Microsoft set allow-all cookies as the default for sites who misconfigure P3P presumably because blocking cookies may have affected those websites. This decision on defaults was widely exploited;  a misconfiguration described on a Microsoft support page was detected down to the exact typo in $2\,756$ sites~\cite{leon2010token}.  Similarly, DNT was adopted without sufficient enforcement from browsers, which does little to improve user privacy beyond shifting the blame to AdTech vendors for not respecting the signal. 

More encouragingly, history also shows privacy advocates can subvert systems with relatively low-effort browser add-ons.  For example, advertising networks expected every user to visit their individual websites to set opt-out cookies~\cite{dixon2007network}.  In reality, the TACO browser extension allowed one individual to maintain and share an updated list of cookies with thousands of users~\cite{soghoian2011thehistory}.  Similarly, the Privacy Bird allegedly helped boost P3P adoption by directly making the user aware of websites' adoption decisions.  These two examples point to the importance of designing privacy enhancing technologies that allow users to send low-effort privacy preference signals.  This becomes especially urgent given the state of the present, to which we now turn.

\subsection{Present} Having surveyed a history in which P3P and DNT were eventually deprecated and NAI membership remains at less than one hundred vendors, our measurements provide an updated picture as of February 2021.  TCF is the dominant signal as the GPC was released as an unofficial draft in October $2020$ and only six websites in the Tranco top 100k now implement it.  Given signals must be adopted by both sender and recipient, we now discuss adoption among each stakeholder.

Websites are arguably the most important stakeholder for the success of TCF since only websites can collect consent signals~\cite{woods2020commodification}.  We discovered \bignum{\tcfAllCountInTrancoHundredK} TCF implementations in the top 100k.  A crude comparison can be drawn with a 2010 sample detecting $19.8$k P3P implementations~\cite{leon2010token}.  Turning to estimates that reference a toplist, TCF is more prevalent among both the top 5k ($\bignum{int(\tcfAllCountInTrancoFiveK/50)}\%$) and top 100k ($\bignum{int(\tcfAllCountInTrancoHundredK/1000)}\%$) than historic P3P measurements ($8\%$~\cite{beatty2007p3p} and $2\%$~\cite{reay2013privacy} respectively).  Such comparisons are limited by changes in the Web and also research methods; P3P adoption studies relied on commercial rankings, whereas we used a top list designed to be stable over time for research purposes. 
This should make our measurement more comparable to future work. %

Turning to adoption among AdTech vendors, vendors were early adopters of TCF and also the first to migrate to \tcftwo{} (see Figure~\ref{fig:tcf-share-timeline}).  By October $2020$, more than $600$ vendors had adopted TCF.  For comparison, just $75$ vendors were offering opt-out cookies in June $2010$ of which only $11$ were also implementing P3P~\cite{leon2010token}.  Although AdTech vendors drafted the TCF specification, adoption was not inevitable given the NAI had no more than $6$ full members from $2001$--$2007$~\cite{dixon2007network}.  Thus, TCF is the first privacy preference signal to achieve widespread adoption among AdTech vendors.

Our results also speak to why websites are adopting TCF.  Numerous pieces of evidence suggest vendors incentivize partner websites to adopt TCF (see Figure~\ref{fig:tcf-bytoplist}, Figure~\ref{fig:tcf-by-tp} and especially Figure~\ref{fig:google-impact}).  An interesting comparision can be made with P3P. Websites embedding more third-party domains are more likely to adopt TCF but less likely to adopt P3P~\cite[p.~292]{cranor2008p3p}.  This supports the common sense intuition that TCF was designed to perpetuate privacy eroding business models.  

More generally, we provide evidence in support of the general finding that private firms deploy economic resources to ensure the adoption of standards~\cite{shapiro1998information}.  Figure~\ref{fig:tcf-bytoplist} shows how Quantcast's free consent management solution supported TCF adoption, particularly among less popular sites.  The role of institutional support is crucial even to open standards, such as the organization of TCP/IP education events~\cite{leiner2009brief} and subsidization of free certificates via Let's Encrypt to support HTTPS adoption~\cite{felt2017measuring}.   
In terms of migrating to updated standards, we show how Quantcast boosted \tcftwo{} adoption by adding prominent deprecation messages into consent dialogs. Thus, Figure~\ref{fig:qc-transition} suggests that IAB policy (\tcfone{} consent strings becoming invalid) led to Quantcast losing customers.

Finally, we can quantify the relative decision volume of users relative to vendors.  Quantcast boast of processing $25$ billion consent signals~\cite{quantcast2020}, whereas we observed just \bignum{\gvlAllPurposeChanges} changes in vendor purposes since $2018$.  This means users have made at least $\bignum{int(25000/\gvlAllPurposeChanges)}$ million times more decisions than vendors since TCF was launched.  At $3.2$s per decision~\cite{hils2020measuring}, this means users have spent at least 2,500 years since $2018$ expressing their privacy preferences through Quantcast dialogs alone. \label{pageref:envelope}

\subsection{Future} \vspace{-2ex} Given this startling time investment in sending TCF signals, it is worth considering what the future holds for pro-privacy signals.  Releasing the GPC specification in an unofficial draft~\cite{gpc2020} over two years after GDPR came into effect and ten months after CCPA provided TCF with a first-mover advantage. However, we have few concerns that privacy aware users will adopt the GPC in the future.  Pro-privacy browsers like Firefox supported the design, additionally the Brave browser\footnote{\url{https://brave.com/global-privacy-control/}} and add-ons like Privacy Badger\footnote{\url{https://www.eff.org/gpc-privacy-badger}} already turn the GPC signal on by default.

We are less optimistic that the intended recipients, namely AdTech vendors, will adopt the GPC signal.  Much like with DNT~\cite{iab2012dnt}, AdTech vendors are likely to claim that on-as-default makes the signal meaningless.  However, privacy advocates can now rely on privacy laws like the CCPA, which was not available when DNT was first adopted by browsers.  

Fighting legal cases to establish a favorable precedent is a likely strategy.  One of the GPC's participating organizations, Brave Brows\-er, has already lodged complaints under the GDPR against rival browsers~\cite{brave2018google}, national regulators~\cite{brave2020govts}, and even the IAB Europe's website~\cite{brave2019iab}.  
We anticipate similar actions under the CCPA, especially given California's attorney general tweeted about the GPC in January 2021\footnote{\url{https://digiday.com/media/why-a-tweet-from-californias-ag-about-a-global-privacy-tool-has-companies-scrambling/}}.
Multiple publishers adopting the same standard and out-sourcing implementation to dominant CMPs creates the potential for auditing at scale~[p.\,10]\cite{hils2020measuring}, as evidenced by an NGO's threat of automated complaints against publishers\footnote{\url{https://noyb.eu/en/noyb-aims-end-cookie-banner-terror-and-issues-more-500-gdpr-complaints}}.

Regulatory interventions may begin to undermine the adoption of TCF.  For example, the Danish regulator ruled that the Danish Meteorological Institute could not claim a legitimate interest in collecting personal data~\cite{danishdpa2020}.  
Possibly preempting such a ruling, the option to declare a legitimate interest in storing and/or accessing information on a device was removed in \tcftwo{} (see Figure~\ref{fig/gvl-transition-storage}).
The case also ruled that opt-out must be as easy as opt-in. 
Many websites collecting TCF signals do not follow this ruling~\cite{nouwens2020dark, hils2020measuring}.   
The leading provider of TCF dialogs distances itself from ambiguity in privacy law~\cite{ceross2018rethinking} by making the design choice a configuration that websites select, with one CMP warning ``with great customizability comes great responsibility''~\cite{hils2020measuring}.
 This indicates that AdTech vendors perceive liability risk related to TCF.
This discussion raises the question of what happens when two signals co-exist.
Whereas standards usually have a definitive winner, such as DVD over DIVX or VHS over Betamax~\cite{shapiro1999art}, GPC and TCF signals can be sent simultaneously because they are defined on different network layers (see Table~\ref{fig:standards-table}).  Encouragingly, one could imagine a future in which browsers exploit control over what is rendered to the user to block dialogs from loading, whereas AdTech cannot stop browsers from sending GPC headers as part of HTTP requests.  Signals co-existing is more troublesome when it comes to interpretation.  A TCF opt-in signal could be sent in an HTTP request with GPC opt-out headers. We leave it to legal scholars and future court cases to ponder which signal has priority.

Arguably this back and forth over privacy preference signals has been a distraction for over $20$ years.  Regardless of the adoption of privacy preference signals, there is little basis to trust that expressed preferences will be respected.  In terms of what we can observe: vendors ignoring the DNT signal was public policy~\cite{iab2012dnt}, P3P was intentionally misconfigured by websites~\cite{leon2010token}, TCF consent signals misreport the user's expressed preferences~\cite{matte2020do}, tracking remains ubiquitous in a post-GDPR world~\cite{sanchez2019} and there is growing evidence firms use dark patterns to manipulate users' expressed preferences~\cite{bosch2016tales, mathur2019dark, narayanan2020dark}.  More fundamentally, there is no way of auditing whether AdTech vendors respect expressed signals.

\vspace{-3ex} 
\section{Conclusion} \vspace{-2ex} \label{section:conclusion}
Privacy preference signals must be adopted by both senders (users) and recipients (AdTech vendors) who have differing requirements.  Vendors want to receive positive consent signals in order to comply with privacy laws, and prefer not to receive negative signals that undermine the vendor's business model.  This reasoning helps to explain why hundreds of vendors adopted TCF~\cite{hils2020measuring, matte2020purposes}, which represents a historical anomaly given vendors reluctance to adopt P3P~\cite{leon2010token}, DNT~\cite{dntimplementors} and NAI opt-out cookies~\cite{dixon2007network}.  Our evidence that vendors were early adopters of \tcftwo{} (Figure~\ref{fig:tcf-share-timeline}) underlies the AdTech vendors' commitment to receiving these signals.

History reveals two approaches to collecting users' privacy preferences that are represented in the signal, namely via the user agent (as in P3P and DNT) or a webpage (as in NAI opt-out).  As with the previous signal designed by AdTech~\cite{dixon2007network}, TCF collects user preferences via dialogs embedded in a web page but this requires adoption among websites. Our results show website adoption varies from 5\% to 12\% across sections of the Tranco top 100k (Figure~\ref{fig:tcf-bytoplist}) and is most prevalent among News \& Entertainment websites (Figure~\ref{fig:tcf-by-category}). We also show that the presence of Google Ads (Figure~\ref{fig:google-impact}) and the number of embedded parties (Figure~\ref{fig:tcf-by-tp}) are both associated with greater TCF adoption rates.

Adoption is further supported by AdTech actors like Quantcast lowering the cost of adopting TCF by providing free dialogs marketed as compliant with GDPR (although legality has been called into question~\cite{nouwens2020dark, matte2020do}).  The increase in adoption following May 2018, which can be seen in Figure~\ref{fig:qc-transition}, shows how AdTech capitalised on the passage of the GDPR.  This means AdTech firms now not only draft the TCF, but also actively manage and configure it.  This market power facilitated the swift transition to \tcftwo{} (see Figure~\ref{fig:ot-transition} and Figure~\ref{fig:qc-transition}),  which is remarkable when contrasted against the time to migrate to HTTPS~\cite{felt2017measuring} or IPv6~\cite{nikkah2016migrating}.  

Thus, our measurements of the present reveal TCF is now the dominant privacy preference signal.  Further, its adoption among \emph{both} senders and recipients is a significant historical development (see Table~\ref{fig:standards-table}).  Adoption among recipients is unsurprising given the working group who designed TCF was controlled by the Interactive Advertising Bureau and contained no privacy advocates.  However, websites appear to have sided with their business partners over users.  Consequently, users are forced to send signals via time consuming dialogs.  Our back-of-the-envelope calculation on p.\,\pageref{pageref:envelope} suggests over two thousand years of user time has been spent on sending TCF consent signals since $2018$.  All stakeholders should ask to what extent the TCF's fine-grained, site-by-site signal clarifying privacy preferences has materially changed how recipients process personal data?  A second question is whether a revised signal would lead to better outcomes, or can the problems only be resolved by the technical constraints of \emph{hard privacy}?

\vspace{-3ex}
\section*{Acknowledgements}
\vspace{-2ex}
We would like to thank Aldo Cortesi for his continuous support and the generous access to the Netograph API and capturing technology.
We thank Anelia Kurteva, Jérémie Bernard Glossi, Dennis Jackson, our shepherd Christo Wilson, and the other anonymous reviewers for many constructive comments. The second author is funded by the European Commission's call H2020-MSCA-IF-2019 under grant number 894700.

\bibliographystyle{unsrt}
\bibliography{paper}

\begin{thebibliography}{10}

\bibitem{leon2010token}
Pedro~Giovanni Leon, Lorrie~Faith Cranor, Aleecia~M McDonald, and Robert
  McGuire.
\newblock Token attempt: {T}he misrepresentation of website privacy policies
  through the misuse of {P3P} compact policy tokens.
\newblock In {\em ACM Workshop on Privacy in the Electronic Society}, pages
  93--104, 2010.

\bibitem{center2000pretty}
{Electronic Privacy Information Center} and {Junkbusters}.
\newblock {Pretty Poor Privacy: An Assessment of {P3P} and Internet Privacy}.
\newblock \url{https://epic.org/reports/prettypoorprivacy.html}, 2000.

\bibitem{tpwg2019wg}
{Tracking Protection Working Group}.
\newblock {WG closed}.
\newblock \url{https://github.com/w3c/dnt/commit/5d85d6c}, 2019.

\bibitem{iab2012dnt}
{Interactive Advertising Bureau}.
\newblock {``Do Not Track'' set to ``On'' by Default in Internet Explorer
  10---IAB Response}.
\newblock
  \url{https://www.iab.com/news/do-not-track-set-to-on-by-default-in-internet-explorer-10iab-response/},
  2012.

\bibitem{dixon2007network}
Pam Dixon.
\newblock {\em {The Network Advertising Initiative: Failing at Consumer
  Protection and at Self-Regulation}}.
\newblock World Privacy Forum, 2007.
\newblock
  \url{http://www.worldprivacyforum.org/wp-content/uploads/2007/11/WPF_NAI_report_Nov2_2007fs.pdf}.

\bibitem{united1998privacy}
Martha~K. Landesberg, Toby~Milgrom Levin, Caroline~G. Curtin, and Ori Lev.
\newblock {\em Privacy online: A Report to Congress}.
\newblock US Federal Trade Commission, 1998.

\bibitem{utz2019uninformed}
Christine Utz, Martin Degeling, Sascha Fahl, Florian Schaub, and Thorsten Holz.
\newblock {(Un)informed Consent: Studying {GDPR} Consent Notices in the Field}.
\newblock In {\em Proceedings of the 2019 {ACM} {SIGSAC} Conference on Computer
  and Communications Security}, CCS '19, pages 973--990. {ACM}, 2019.

\bibitem{matte2020do}
C{\'{e}}lestin Matte, Nataliia Bielova, and Cristiana Santos.
\newblock {Do Cookie Banners Respect my Choice? Measuring Legal Compliance of
  Banners from {IAB} Europe's Transparency and Consent Framework}.
\newblock In {\em {IEEE} Symposium on Security and Privacy}, pages 791--809.
  {IEEE}, 2020.

\bibitem{nouwens2020dark}
Midas Nouwens, Ilaria Liccardi, Michael Veale, David Karger, and Lalana Kagal.
\newblock {Dark Patterns after the GDPR: Scraping Consent Pop-Ups and
  Demonstrating Their Influence}.
\newblock In {\em Proceedings of the 2020 CHI Conference on Human Factors in
  Computing Systems}, CHI '20. {ACM}, 2020.

\bibitem{machuletz2020}
Dominique Machuletz and Rainer B{\"o}hme.
\newblock {Multiple Purposes, Multiple Problems: {A} User Study of Consent
  Dialogs after {GDPR}}.
\newblock {\em Proceedings on Privacy Enhancing Technologies}, (2):481--498,
  2020.

\bibitem{habib2020scavenger}
Hana Habib, Sarah Pearman, Jiamin Wang, Yixin Zou, Alessandro Acquisti,
  Lorrie~Faith Cranor, Norman Sadeh, and Florian Schaub.
\newblock {"It's a Scavenger Hunt": Usability of Websites' Opt-Out and Data
  Deletion Choices}.
\newblock In {\em Proceedings of the 2020 CHI Conference on Human Factors in
  Computing Systems}, CHI '20. ACM, 2020.

\bibitem{oconnor2020unclear}
Sean O'Connor, Ryan Nurwono, and Eleanor Birrell.
\newblock {(Un)clear and (In)conspicuous: The right to opt-out of sale under
  CCPA}, 2020.

\bibitem{hils2020measuring}
Maximilian Hils, Daniel~W Woods, and Rainer B{\"{o}}hme.
\newblock {Measuring the Emergence of Consent Management on the Web}.
\newblock In {\em Proceedings of the Internet Measurement Conference 2020}, IMC
  '20. {ACM}, 2020.

\bibitem{woods2020commodification}
Daniel~W Woods and Rainer B{\"{o}}hme.
\newblock The commodification of consent.
\newblock In {\em 20th Annual Workshop on the Economics of Information
  Security, {WEIS}}, 2020.

\bibitem{acar2014theweb}
Gunes Acar, Christian Eubank, Steven Englehardt, Marc Juarez, Arvind Narayanan,
  and Claudia Diaz.
\newblock {The Web Never Forgets: Persistent Tracking Mechanisms in the Wild}.
\newblock In {\em Proceedings of the 2014 ACM SIGSAC Conference on Computer and
  Communications Security}, CCS '14, pages 674--689. ACM, 2014.

\bibitem{englehardt2015cookies}
Steven Englehardt, Dillon Reisman, Christian Eubank, Peter Zimmerman, Jonathan
  Mayer, Arvind Narayanan, and Edward~W. Felten.
\newblock {Cookies That Give You Away: The Surveillance Implications of Web
  Tracking}.
\newblock In {\em Proceedings of the 24th International Conference on World
  Wide Web}, WWW '15, pages 289--299, Republic and Canton of Geneva, CHE, 2015.
  International World Wide Web Conferences Steering Committee.

\bibitem{englehardt2016online}
Steven Englehardt and Arvind Narayanan.
\newblock {Online Tracking: A 1-Million-Site Measurement and Analysis}.
\newblock In {\em Proceedings of the 2016 ACM SIGSAC Conference on Computer and
  Communications Security}, CCS '16, pages 1388--1401. ACM, 2016.

\bibitem{laperdrix2020browser}
Pierre Laperdrix, Nataliia Bielova, Benoit Baudry, and Gildas Avoine.
\newblock {Browser Fingerprinting: A Survey}.
\newblock {\em ACM Trans. Web}, 14(2), April 2020.

\bibitem{bujlow2017asurvey}
T.~{Bujlow}, V.~{Carela-Espa{\~n}ol}, J.~{Sol{\'e}-Pareta}, and
  P.~{Barlet-Ros}.
\newblock {A Survey on Web Tracking: Mechanisms, Implications, and Defenses}.
\newblock {\em Proceedings of the IEEE}, 105(8):1476--1510, 2017.

\bibitem{byers2003automated}
Simon Byers, Lorrie~Faith Cranor, and David Kormann.
\newblock Automated analysis of {P3P}-enabled web sites.
\newblock In {\em Proceedings of the 5th International Conference on Electronic
  Commerce}, pages 326--338, 2003.

\bibitem{beatty2007p3p}
Patricia Beatty, Ian Reay, Scott Dick, and James Miller.
\newblock {P3P} adoption on e-commerce web sites: a survey and analysis.
\newblock {\em IEEE Internet Computing}, 11(2):65--71, 2007.

\bibitem{reay2013privacy}
Ian Reay, Patricia Beatty, Scott Dick, and James Miller.
\newblock Privacy policies and national culture on the internet.
\newblock {\em Information Systems Frontiers}, 15(2):279--292, 2013.

\bibitem{richmond2010loophole}
Riva Richmond.
\newblock A loophole big enough for a cookie to fit through.
\newblock {\em New York Times}, 2010.
\newblock \url{https://nyti.ms/2mDvTBQ}.

\bibitem{cranor2002use}
Lorrie~Faith Cranor, Manjula Arjula, and Praveen Guduru.
\newblock Use of a {P3P} user agent by early adopters.
\newblock In {\em Proceedings of the 2002 ACM Workshop on Privacy in the
  Electronic Society}, pages 1--10, 2002.

\bibitem{w3c2011dnt}
{World Wide Web Consortium}.
\newblock {Tracking Protection Working Group}.
\newblock \url{https://www.w3.org/2011/tracking-protection/}, 2011.

\bibitem{wsj2012dnt}
Julia Angwin.
\newblock {Microsoft's ``Do Not Track'' Move Angers Advertising Industry}.
\newblock \url{https://www.wsj.com/articles/BL-DGB-24506}, 2012.

\bibitem{chrome2012longer}
{Chrome Blog}.
\newblock {Longer battery life and easier website permissions}.
\newblock
  \url{https://chrome.googleblog.com/2012/11/longer-battery-life-and-easier-website.html},
  2012.

\bibitem{dntimplementors}
Future of~Privacy~Forum.
\newblock {Companies that have implemented Do Not Track}.
\newblock \url{https://allaboutdnt.com/companies/}, 2020.

\bibitem{fowler2013dnt}
Alex Fowler.
\newblock { Mozilla's new Do Not Track dashboard: Firefox users continue to
  seek out and enable DNT}.
\newblock
  \url{https://blog.mozilla.org/netpolicy/2013/05/03/mozillas-new-do-not-track-dashboard-firefox-users-continue-to-seek-out-and-enable-dnt/},
  2013.

\bibitem{gpc2020}
Robin Berjon, Sebastian Zimmeck, Ashkan Soltani, David Harbage, and Peter
  Synder.
\newblock {Global Privacy Control (GPC) Unofficial Draft 15 October 2020}.
\newblock \url{https://globalprivacycontrol.github.io/gpc-spec/}, 2020.

\bibitem{iabeurope2020what}
{IAB Europe}.
\newblock {What is the Transparency and Consent Framework (TCF)?}
\newblock \url{https://iabeurope.eu/transparency-consent-framework/}, 2020.

\bibitem{mayer2012third}
J.~R. {Mayer} and J.~C. {Mitchell}.
\newblock Third-party web tracking: Policy and technology.
\newblock In {\em 2012 {IEEE} Symposium on Security and Privacy}, pages
  413--427. {IEEE}, 2012.

\bibitem{krishnamurthy2009leakage}
Balachander Krishnamurthy and Craig~E Wills.
\newblock On the leakage of personally identifiable information via online
  social networks.
\newblock In {\em Proceedings of the 2nd {ACM} workshop on online social
  networks}, pages 7--12, 2009.

\bibitem{acar2020dataexfiltration}
Gunes Acar, Steven Englehardt, and Arvind Narayanan.
\newblock No boundaries: data exfiltration by third parties embedded on web
  pages.
\newblock {\em Proceedings on Privacy Enhancing Technologies}, 2020(4):220 --
  238, 2020.

\bibitem{farooqi2020canarytrap}
Shehroze Farooqi, Maaz Musa, Zubair Shafiq, and Fareed Zaffar.
\newblock Canarytrap: Detecting data misuse by third-party apps on online
  social networks.
\newblock {\em Proceedings on Privacy Enhancing Technologies}, 2020(4):336 --
  354, 2020.

\bibitem{reyes2018think}
Irwin Reyes, Primal Wijesekera, Joel Reardon, Amit Elazari~Bar On, Abbas
  Razaghpanah, Narseo Vallina-Rodriguez, and Serge Egelman.
\newblock {``Won't Somebody Think of the Children?'' Examining COPPA Compliance
  at Scale}.
\newblock {\em Proceedings on Privacy Enhancing Technologies}, 2018(3):63 --
  83, 2018.

\bibitem{saleem2020sok}
Hamza Saleem and Muhammad Naveed.
\newblock {SoK: Anatomy of Data Breaches}.
\newblock {\em Proceedings on Privacy Enhancing Technologies}, 2020(4):153 --
  174, 2020.

\bibitem{henri2020multihoming}
S{\'e}bastien Henri, Gines Garcia-Aviles, Pablo Serrano, Albert Banchs, and
  Patrick Thiran.
\newblock {Protecting against Website Fingerprinting with Multihoming}.
\newblock {\em Proceedings on Privacy Enhancing Technologies}, 2020(2):89 --
  110, 01 Apr. 2020.

\bibitem{mazmudar2020mitigator}
Miti Mazmudar and Ian Goldberg.
\newblock {Mitigator: Privacy policy compliance using trusted hardware}.
\newblock {\em Proceedings on Privacy Enhancing Technologies}, 2020(3):204 --
  221, 2020.

\bibitem{trevisan2019years}
Martino Trevisan, Stefano Traverso, Eleonora Bassi, and Marco Mellia.
\newblock {4 Years of EU Cookie Law: Results and Lessons Learned}.
\newblock {\em Proceedings on Privacy Enhancing Technologies}, 2019(2):126 --
  145, 2019.

\bibitem{WB2021-SoK}
Daniel~W. Woods and Rainer B{\"o}hme.
\newblock {SoK}: {Q}uantifying cyber risk.
\newblock In {\em {IEEE} Symposium on Security and Privacy}, May 2021.

\bibitem{shipp2020how}
Laura Shipp and Jorge Blasco.
\newblock {How private is your period?: A systematic analysis of menstrual app
  privacy policies}.
\newblock {\em Proceedings on Privacy Enhancing Technologies}, 2020(4):491 --
  510, 2020.

\bibitem{amos2020privacy}
Ryan Amos, Gunes Acar, Elena Lucherini, Mihir Kshirsagar, Arvind Narayanan, and
  Jonathan Mayer.
\newblock {Privacy Policies over Time: Curation and Analysis of a
  Million-Document Dataset}.
\newblock {\em arXiv preprint arXiv:2008.09159}, 2020.

\bibitem{degeling2019}
Martin Degeling, Christine Utz, Christopher Lentzsch, Henry Hosseini, Florian
  Schaub, and Thorsten Holz.
\newblock {We Value Your Privacy ... Now Take Some Cookies: Measuring the
  GDPR's Impact on Web Privacy}.
\newblock In {\em 26th Annual Network and Distributed System Security
  Symposium}, NDSS '19. The Internet Society, 2019.

\bibitem{linden2020landscape}
Thomas Linden, Rishabh Khandelwal, Hamza Harkous, and Kassem Fawaz.
\newblock {The Privacy Policy Landscape After the GDPR}.
\newblock {\em Proceedings on Privacy Enhancing Technologies}, 2020(1):47 --
  64, 01 Jan. 2020.

\bibitem{olson2005study}
Judith~S Olson, Jonathan Grudin, and Eric Horvitz.
\newblock A study of preferences for sharing and privacy.
\newblock In {\em CHI'05 extended abstracts on Human factors in Computing
  Systems}, pages 1985--1988, 2005.

\bibitem{ackerman1999privacy}
Mark~S Ackerman, Lorrie~Faith Cranor, and Joseph Reagle.
\newblock Privacy in e-commerce: examining user scenarios and privacy
  preferences.
\newblock In {\em Proceedings of the 1st ACM Conference on Electronic
  commerce}, pages 1--8, 1999.

\bibitem{weinshel2019oh}
Ben Weinshel, Miranda Wei, Mainack Mondal, Euirim Choi, Shawn Shan, Claire
  Dolin, Michelle~L. Mazurek, and Blase Ur.
\newblock {Oh, the Places You've Been! User Reactions to Longitudinal
  Transparency About Third-Party Web Tracking and Inferencing}.
\newblock In {\em Proceedings of the 2019 ACM SIGSAC Conference on Computer and
  Communications Security}, CCS '19, pages 149--166. ACM, 2019.

\bibitem{spiekerman2001eprivacy}
Sarah Spiekermann, Jens Grossklags, and Bettina Berendt.
\newblock {E-Privacy in 2nd Generation E-Commerce: Privacy Preferences versus
  Actual Behavior}.
\newblock In {\em Proceedings of the 3rd ACM Conference on Electronic
  Commerce}, EC '01, pages 38--47. ACM, 2001.

\bibitem{barth2017privacy}
Susanne Barth and Menno~DT De~Jong.
\newblock The privacy paradox -- {I}nvestigating discrepancies between
  expressed privacy concerns and actual online behavior -- {A} systematic
  literature review.
\newblock {\em Telematics and informatics}, 34(7):1038--1058, 2017.

\bibitem{gerber2018explaining}
Nina Gerber, Paul Gerber, and Melanie Volkamer.
\newblock Explaining the privacy paradox: A systematic review of literature
  investigating privacy attitude and behavior.
\newblock {\em Computers \& Security}, 77:226--261, 2018.

\bibitem{cranor2003p3p}
Lorrie~Faith Cranor.
\newblock {P3P}: Making privacy policies more useful.
\newblock {\em IEEE Security \& Privacy}, 1(6):50--55, 2003.

\bibitem{agrawal2005xpref}
Rakesh Agrawal, Jerry Kiernan, Ramakrishnan Srikant, and Yirong Xu.
\newblock {XPref: a preference language for {P3P}}.
\newblock {\em Computer Networks}, 48(5):809 -- 827, 2005.
\newblock Web Security.

\bibitem{iyilade2014p2u}
Johnson Iyilade and Julita Vassileva.
\newblock {P2U: a privacy policy specification language for secondary data
  sharing and usage}.
\newblock In {\em 2014 IEEE Security and Privacy Workshops}, pages 18--22.
  IEEE, 2014.

\bibitem{yang2012language}
Jean Yang, Kuat Yessenov, and Armando Solar-Lezama.
\newblock A language for automatically enforcing privacy policies.
\newblock {\em ACM SIGPLAN Notices}, 47(1):85--96, 2012.

\bibitem{monir2015appl}
Monir Azraoui, Kaoutar Elkhiyaoui, Melek {\"O}nen, Karin Bernsmed,
  Anderson~Santana De~Oliveira, and Jakub Sendor.
\newblock {A-PPL: An Accountability Policy Language}.
\newblock In {\em Data Privacy Management, Autonomous Spontaneous Security, and
  Security Assurance}, pages 319--326, Cham, 2015. Springer.

\bibitem{kagal2008using}
Lalana Kagal, Chris Hanson, and Daniel Weitzner.
\newblock Using dependency tracking to provide explanations for policy
  management.
\newblock In {\em 2008 IEEE Workshop on Policies for Distributed Systems and
  Networks}, pages 54--61. IEEE, 2008.

\bibitem{kumaraguru2007survey}
Ponnurangam Kumaraguru, Lorrie Cranor, Jorge Lobo, and Seraphin Calo.
\newblock A survey of privacy policy languages.
\newblock In {\em Workshop on Usable IT Security Management (USM 07):
  Proceedings of the 3rd Symposium on Usable Privacy and Security, ACM}, 2007.

\bibitem{zhao2016privacylanguages}
Jun Zhao, Reuben Binns, Max Van~Kleek, and Nigel Shadbolt.
\newblock Privacy languages: Are we there yet to enable user controls?
\newblock In {\em Proceedings of the 25th International Conference Companion on
  World Wide Web}, WWW '16 Companion, pages 799--806. International World Wide
  Web Conferences Steering Committee, 2016.

\bibitem{kasem2015security}
Saffija Kasem{-}Madani and Michael Meier.
\newblock Security and privacy policy languages: {A} survey, categorization and
  gap identification.
\newblock {\em CoRR}, abs/1512.00201, 2015.

\bibitem{morel2020}
Victor Morel and Ra{\'{u}}l Pardo.
\newblock {SoK}: Three facets of privacy policies.
\newblock In {\em WPES'20: Proceedings of the 19th Workshop on Privacy in the
  Electronic Society, Virtual Event, USA, November 9, 2020}, pages 41--56.
  {ACM}, 2020.

\bibitem{cranor2008p3p}
Lorrie~Faith Cranor, Serge Egelman, Steve Sheng, Aleecia~M McDonald, and Abdur
  Chowdhury.
\newblock {P3P} deployment on websites.
\newblock {\em Electronic Commerce Research and Applications}, 7(3):274--293,
  2008.

\bibitem{reay2009}
Ian Reay, Scott Dick, and James Miller.
\newblock {An analysis of privacy signals on the World Wide Web: Past, present
  and future}.
\newblock {\em Inf. Sci.}, 179(8):1102--1115, 2009.

\bibitem{matte2020purposes}
C{\'e}lestin Matte, Cristiana Santos, and Nataliia Bielova.
\newblock Purposes in {IAB Europe's TCF}: which legal basis and how are they
  used by advertisers?
\newblock In {\em Annual Privacy Forum}, 2020.

\bibitem{lai2006internet}
Yee-Lin Lai and Kai-Lung Hui.
\newblock Internet opt-in and opt-out: Investigating the roles of frames,
  defaults and privacy concerns.
\newblock In {\em Proceedings of the 2006 ACM SIGMIS CPR Conference on Computer
  Personnel Research}, SIGMIS CPR '06, pages 253--263. {ACM}, 2006.

\bibitem{boehme2010}
Rainer B{\"{o}}hme and Stefan K{\"{o}}psell.
\newblock Trained to accept? {A} field experiment on consent dialogs.
\newblock In {\em Proceedings of the SIGCHI Conference on Human Factors in
  Computing Systems}, CHI '10, pages 2403--2406. {ACM}, 2010.

\bibitem{adjerid2013sleights}
Idris Adjerid, Alessandro Acquisti, Laura Brandimarte, and George Loewenstein.
\newblock Sleights of privacy: Framing, disclosures, and the limits of
  transparency.
\newblock In {\em Proceedings of the Ninth Symposium on Usable Privacy and
  Security}, SOUPS '13. {ACM}, 2013.

\bibitem{leiner2009brief}
Barry~M Leiner, Vinton~G Cerf, David~D Clark, Robert~E Kahn, Leonard Kleinrock,
  Daniel~C Lynch, Jon Postel, Larry~G Roberts, and Stephen Wolff.
\newblock A brief history of the internet.
\newblock {\em ACM SIGCOMM Computer Communication Review}, 39(5):22--31, 2009.

\bibitem{nikkhah2017statistical}
Mehdi Nikkhah, Aman Mangal, Constantine Dovrolis, and Roch Gu{\'e}rin.
\newblock A statistical exploration of protocol adoption.
\newblock {\em IEEE/ACM Transactions on Networking}, 25(5):2858--2871, 2017.

\bibitem{czyz2014measuring}
Jakub Czyz, Mark Allman, Jing Zhang, Scott Iekel-Johnson, Eric Osterweil, and
  Michael Bailey.
\newblock Measuring {IPv6} adoption.
\newblock {\em SIGCOMM Comput. Commun. Rev.}, 44(4):87--98, August 2014.

\bibitem{wang2018extending}
Xuequn Wang and Sebastian Zander.
\newblock Extending the model of internet standards adoption: A cross-country
  comparison of {IPv6} adoption.
\newblock {\em Information \& Management}, 55(4):450 -- 460, 2018.

\bibitem{nikkah2016migrating}
M.~{Nikkhah} and R.~{Gu{\'e}rin}.
\newblock {Migrating the Internet to IPv6: An Exploration of the When and Why}.
\newblock {\em IEEE/ACM Transactions on Networking}, 24(4):2291--2304, 2016.

\bibitem{holz2011thessl}
Ralph Holz, Lothar Braun, Nils Kammenhuber, and Georg Carle.
\newblock {The SSL Landscape: A Thorough Analysis of the x.509 PKI Using Active
  and Passive Measurements}.
\newblock In {\em Proceedings of the 2011 ACM SIGCOMM Conference on Internet
  Measurement Conference}, IMC '11, pages 427--444. ACM, 2011.

\bibitem{ozment2006bootstrapping}
Andy Ozment and Stuart~E Schechter.
\newblock Bootstrapping the adoption of internet security protocols.
\newblock In {\em 5th Annual Workshop on the Economics of Information Security,
  {WEIS}}, 2006.

\bibitem{felt2017measuring}
Adrienne~Porter Felt, Richard Barnes, April King, Chris Palmer, Chris Bentzel,
  and Parisa Tabriz.
\newblock Measuring {HTTPS} adoption on the web.
\newblock In {\em Proceedings of the {USENIX} Security Symposium
  ({USENIX}~Security~17)}, pages 1323--1338, 2017.

\bibitem{pochat2019}
Victor~Le Pochat, Tom van Goethem, Samaneh Tajalizadehkhoob, Maciej Korczynski,
  and Wouter Joosen.
\newblock Tranco: {A} research-oriented top sites ranking hardened against
  manipulation.
\newblock In {\em 26th Annual Network and Distributed System Security
  Symposium}, NDSS '19. The Internet Society, 2019.

\bibitem{rulespace}
Symantec.
\newblock Symantec {RuleSpace}: {URL} categorization database, 2020.

\bibitem{sanchez2019}
Iskander Sanchez-Rola, Matteo Dell'Amico, Platon Kotzias, Davide Balzarotti,
  Leyla Bilge, Pierre-Antoine Vervier, and Igor Santos.
\newblock {Can I Opt Out Yet? GDPR and the Global Illusion of Cookie Control}.
\newblock In {\em Proceedings of the 2019 ACM Asia Conference on Computer and
  Communications Security}, Asia CCS '19, pages 340--351. {ACM}, 2019.

\bibitem{vallina2020}
Pelayo Vallina, Victor Le~Pochat, {\'A}lvaro Feal, Marius Paraschiv, Julien
  Gamba, Tim Burke, Oliver Hohlfeld, Juan Tapiador, and Narseo
  Vallina-Rodriguez.
\newblock {Mis-shapes, Mistakes, Misfits: An Analysis of Domain Classification
  Services}.
\newblock In {\em Proceedings of the Internet Measurement Conference 2020}, IMC
  '20. {ACM}, 2020.

\bibitem{publicsuffix}
Mozilla Foundation.
\newblock Public suffix list.
\newblock \url{https://publicsuffix.org/}, 2007--2020.

\bibitem{dimova2021cname}
Yana Dimova, Gunes Acar, Lukasz Olejnik, Wouter Joosen, and Tom van Goethem.
\newblock {The {CNAME} of the Game: Large-scale Analysis of DNS-based Tracking
  Evasion}.
\newblock {\em Proceedings on Privacy Enhancing Technologies}, 2021.

\bibitem{inside2013digital}
{Inside Privacy}.
\newblock {Digital Advertising Alliance Leaves Do Not Track Group}.
\newblock
  \url{https://www.insideprivacy.com/advertising-marketing/digital-advertising-alliance-leaves-do-not-track-group-2/},
  2013.

\bibitem{iabtechlab2020global}
{IAB Tech Lab}.
\newblock {Global Privacy Working Group}.
\newblock
  \url{https://iabtechlab.com/working-groups/global-privacy-working-group/},
  2011.

\bibitem{russell2006rough}
Andrew~L Russell.
\newblock {`Rough consensus and running code'} and the {Internet-OSI} standards
  war.
\newblock {\em IEEE Annals of the History of Computing}, 28(3):48--61, 2006.

\bibitem{soghoian2011thehistory}
Christopher Soghoian.
\newblock {The History of the Do Not Track Header}.
\newblock
  \url{http://paranoia.dubfire.net/2011/01/history-of-do-not-track-header.html},
  2011.

\bibitem{shapiro1998information}
Carl Shapiro, Shapiro Carl, Hal~R Varian, et~al.
\newblock {\em Information rules: a strategic guide to the network economy}.
\newblock Harvard Business Press, 1998.

\bibitem{quantcast2020}
Kochava Inc.
\newblock {Quantcast and Kochava Partnership Delivers Combined Web and Mobile
  App Solution for CCPA}.
\newblock
  \url{https://www.businesswire.com/news/home/20200207005054/en/Quantcast-and-Kochava-Partnership-Delivers-Combined-Web-and-Mobile-App-Solution-for-CCPA},
  2018.

\bibitem{brave2018google}
Johnny Ryan.
\newblock {Regulatory complaint concerning massive, web-wide data breach by
  Google and other ``ad tech'' companies under Europe's GDPR}.
\newblock \url{https://brave.com/adtech-data-breach-complaint/}, 2018.

\bibitem{brave2020govts}
Natasha Lomas.
\newblock {Brave Accueses European governments of GDPR resourcing failure}.
\newblock
  \url{https://techcrunch.com/2020/04/27/brave-accuses-european-governments-of-gdpr-resourcing-failure/},
  2020.

\bibitem{brave2019iab}
Johnny Ryan.
\newblock {Formal GDPR complaint against IAB Europe's ``cookie wall'' and GDPR
  consent guidance}.
\newblock \url{https://brave.com/iab-cookie-wall/}, 2019.

\bibitem{danishdpa2020}
Tue Goldschmieding.
\newblock {New important decision on cookies from the Danish Data Protection
  Agency}.
\newblock
  \url{https://gorrissenfederspiel.com/en/knowledge/news/new-important-decision-on-cookies-from-the-danish-data-protection-agency},
  2020.

\bibitem{ceross2018rethinking}
Aaron Ceross and Andrew Simpson.
\newblock {Rethinking the Proposition of Privacy Engineering}.
\newblock In {\em Proceedings of the New Security Paradigms Workshop}, NSPW
  '18, pages 89--102. ACM, 2018.

\bibitem{shapiro1999art}
Carl Shapiro and Hal~R Varian.
\newblock The art of standards wars.
\newblock {\em California Management Review}, 41(2):8--32, 1999.

\bibitem{bosch2016tales}
Christoph B{\"o}sch, Benjamin Erb, Frank Kargl, Henning Kopp, and Stefan
  Pfattheicher.
\newblock Tales from the dark side: Privacy dark strategies and privacy dark
  patterns.
\newblock {\em Proceedings on Privacy Enhancing Technologies},
  2016(4):237--254, 2016.

\bibitem{mathur2019dark}
Arunesh Mathur, Gunes Acar, Michael~J Friedman, Elena Lucherini, Jonathan
  Mayer, Marshini Chetty, and Arvind Narayanan.
\newblock Dark patterns at scale: Findings from a crawl of 11k shopping
  websites.
\newblock {\em Proceedings of the ACM on Human-Computer Interaction},
  3(CSCW):1--32, 2019.

\bibitem{narayanan2020dark}
Arvind Narayanan, Arunesh Mathur, Marshini Chetty, and Mihir Kshirsagar.
\newblock {Dark Patterns: Past, Present, and Future}.
\newblock {\em {ACM} Queue}, 18(2):67--92, 2020.

\end{thebibliography}

\appendix
\counterwithin{figure}{section}
\counterwithin{table}{section}
\section{Appendix}

	\ifstandalone\begin{table}\else\begin{table*}\fi
		\caption{Regression Coefficients for \tcftwox{} Adoption} 
		\label{fig/regression} 
		\small	

\begin{tabular}{@{\extracolsep{5pt}}lccc} 
\\[-1.8ex]\hline 
\hline \\[-1.8ex] 
 & \multicolumn{3}{c}{\textit{Dependent variable:}} \\ 
\cline{2-4} 
\\[-1.8ex] & \multicolumn{3}{c}{\tcftwox{} Adoption} \\ 
\\[-1.8ex] & (1) & (2) & (3)\\ 
\hline \\[-1.8ex] 
 Google Ads & 0.150$^{***}$ & 0.151$^{***}$ & 3.502$^{***}$ \\ 
  & (0.043) & (0.044) & (0.140) \\ 
  & & & \\ 
 log(\# contacted SLDs) & 0.765$^{***}$ & 0.596$^{***}$ & 1.787$^{***}$ \\ 
  & (0.020) & (0.020) & (0.053) \\ 
  & & & \\ 
 Category: Business &  & $-$0.436$^{***}$ & $-$0.430$^{***}$ \\ 
  &  & (0.055) & (0.055) \\ 
  & & & \\ 
 Category: Education &  & $-$1.384$^{***}$ & $-$1.385$^{***}$ \\ 
  &  & (0.098) & (0.098) \\ 
  & & & \\ 
 Category: Government &  & $-$2.479$^{***}$ & $-$2.506$^{***}$ \\ 
  &  & (0.303) & (0.304) \\ 
  & & & \\ 
 Category: News \& Entertainment &  & 0.954$^{***}$ & 0.994$^{***}$ \\ 
  &  & (0.031) & (0.031) \\ 
  & & & \\ 
 Category: Shopping &  & $-$0.885$^{***}$ & $-$0.826$^{***}$ \\ 
  &  & (0.067) & (0.067) \\ 
  & & & \\ 
 Category: Technology &  & $-$0.487$^{***}$ & $-$0.469$^{***}$ \\ 
  &  & (0.055) & (0.055) \\ 
  & & & \\ 
 Google Ads * log(\# contacted SLDs) &  &  & $-$1.517$^{***}$ \\ 
  &  &  & (0.058) \\ 
  & & & \\ 
 Constant & $-$4.614$^{***}$ & $-$4.189$^{***}$ & $-$6.558$^{***}$ \\ 
  & (0.045) & (0.048) & (0.121) \\ 
  & & & \\ 
\hline \\[-1.8ex] 
Observations & 92,001 & 82,326 & 82,326 \\ 
McFadden's Pseudo-$R^2$ & 0.08 & 0.13 & 0.14 \\

\hline 
\hline \\[-1.8ex] 
\textit{Note:}  & \multicolumn{3}{r}{$^{*}$p$<$0.1; $^{**}$p$<$0.05; $^{***}$p$<$0.01} \\ 
\end{tabular} 

	\ifstandalone\end{table}\else\end{table*}\fi

	\ifstandalone\begin{table}\else\begin{table*}\fi
		\caption{Summary Statistics} 
		\label{fig/summary-statistics}
		\small
		\begin{tabular}{lcccc}
			\toprule
			Variable & N & Min & Mean & Max\\
			\midrule
			\tcftwox{} Adoption & 92,475 & %
				0 & 0.072 & 1\\
			Google Ads & 92,538 & %
				0 & 0.570 & 1\\
			log(\# contacted SLDs) & 92,538 & %
				0 & 2.189 & 5.004 \\
			Category: Business & 88,269 & %
				0 & 0.114 & 1 \\
			Category: Education & 88,269 & %
				0 & 0.078 & 1 \\
			Category: Government & 88,269 & %
				0 & 0.034 & 1 \\
			Category: News \& Entertainment & 88,269 & %
				0 & 0.210 & 1 \\
			Category: Shopping & 88,269 & %
				0 & 0.086 & 1 \\
			Category: Technology & 88,269 & %
				0 & 0.132 & 1 \\
			\bottomrule
		\end{tabular}
	\ifstandalone\end{table}\else\end{table*}\fi

\begin{figure*}[b!]
	\centering
	\fbox{\includegraphics[width=.7\linewidth]{\paperroot/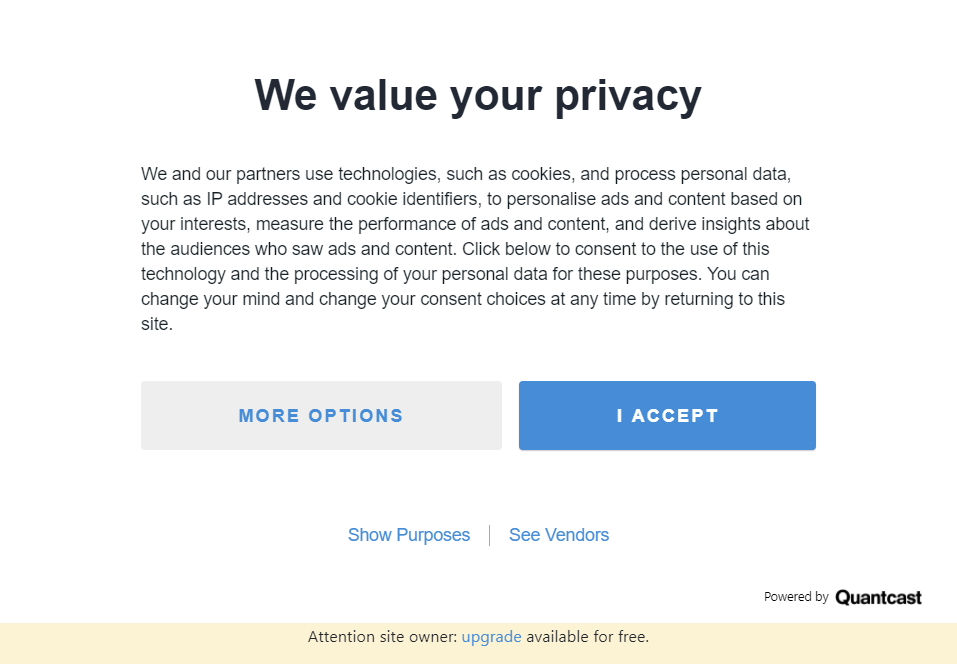}}
	\caption{Starting August 5th 2020, Quantcast added a prominent deprecation message at the bottom of  all its customers' \tcfonex{} consent dialogs, prompting them to switch to \tcftwo{}.}
	\label{fig:qc-attention-site-owner}
\end{figure*}

\end{document}